\documentclass[letterpaper,11pt]{article}
\usepackage{amsmath,mathrsfs,amssymb,amsfonts,amsthm,mathtools}
\usepackage{physics,color}
\usepackage{subcaption}
\usepackage{tabularx}
\usepackage{multirow}
\usepackage{booktabs}
\usepackage[table]{xcolor}
\usepackage{tikz,tikz-cd}
\usetikzlibrary{shapes.geometric, arrows, positioning}
\usepackage{empheq}
\usepackage{bm}

\newcommand*\widefbox[1]{\fbox{\hspace{2em}#1\hspace{2em}}}
\pdfoutput=1
\usepackage{jheppub}
\usepackage[T1]{fontenc}
\usepackage{amsmath} 
\usepackage{mathrsfs} 
\usepackage{xfp} 
\usepackage[outline]{contour} 
\usetikzlibrary{decorations.markings,decorations.pathmorphing}
\usetikzlibrary{angles,quotes} 
\usetikzlibrary{arrows.meta} 
\contourlength{1.4pt}

\tikzset{>=latex} 
\colorlet{myred}{red!80!black}
\colorlet{myblue}{blue!80!black}
\colorlet{mygreen}{green!80!black}
\colorlet{mydarkred}{red!50!black}
\colorlet{mydarkblue}{blue!50!black}
\colorlet{mylightblue}{mydarkblue!6}
\colorlet{mypurple}{blue!40!red!80!black}
\colorlet{mydarkpurple}{blue!40!red!50!black}
\colorlet{mylightpurple}{mydarkpurple!80!red!6}
\colorlet{myorange}{orange!40!yellow!95!black}
\tikzstyle{cone}=[mydarkblue,line width=0.2,top color=blue!60!black!30,
                  bottom color=blue!60!black!50!red!30,shading angle=60,fill opacity=0.9]
\tikzstyle{cone back}=[mydarkblue,line width=0.1,dash pattern=on 1pt off 1pt]
\tikzstyle{world line}=[myblue!60,line width=0.4]
\tikzstyle{world line t}=[mypurple!60,line width=0.4]
\tikzstyle{particle}=[mygreen,line width=0.5]
\tikzstyle{photon}=[-{Latex[length=4,width=3]},myorange,line width=0.4,decorate,
                    decoration={snake,amplitude=0.9,segment length=4,post length=3.8}]
\tikzstyle{singularity}=[myred,line width=0.6,decorate,
                         decoration={zigzag,amplitude=2,segment length=6.17}]
\tikzset{declare function={%
  penrose(\x,\c)  = {\fpeval{2/pi*atan( (sqrt((1+tan(\x)^2)^2+4*\c*\c*tan(\x)^2)-1-tan(\x)^2) /(2*\c*tan(\x)^2) )}};%
  penroseu(\x,\t) = {\fpeval{atan(\x+\t)/pi+atan(\x-\t)/pi}};%
  penrosev(\x,\t) = {\fpeval{atan(\x+\t)/pi-atan(\x-\t)/pi}};%
  kruskal(\x,\c)  = {\fpeval{asin( \c*sin(2*\x) )*2/pi}};
}}

\def\Nsamples{40} 

\newcommand{\nn}{\nonumber}

\hypersetup{colorlinks=true,linkcolor=blue,citecolor=blue,urlcolor=blue}

\title{Exponentially accelerated mirrors as a physical realization of the kappa plane-wave vacuum}

\author[a]{Arash Azizi}
\affiliation[a]{{\it The Institute for Quantum Science and Engineering,
Texas A\&M University,\\ College Station, TX 77843, U.S.A.}}
\emailAdd{sazizi@tamu.edu}

\abstract{
The kappa plane-wave vacuum is a family of kinematically defined quantum states whose thermal properties are well understood, but whose physical origin has remained obscure. In this paper we provide a concrete dynamical realization of this vacuum, showing that it is physically and operationally equivalent to the quantum state produced on future null infinity by a mirror following the Carlitz--Willey (CW) trajectory. The equivalence is established through a three-pronged analysis: we demonstrate that the two constructions share identical Bogoliubov squeeze parameters, identical nonlocal thermal kernels in their Wightman functions, and identical Planckian responses of an Unruh--DeWitt detector. This result anchors an abstract kinematic construction in a well-understood dynamical model, identifying the parameter $\kappa$ with the physical  scale that governs the Carlitz--Willey trajectory. In the final part of the paper we characterize, within the moving-mirror framework, the complete class of mirror trajectories that reproduce the same asymptotic thermal kernel on $\mathscr I^+_R$, and show that only the purely exponential CW trajectory generates a constant, stationary flux.
}

\begin{document}
\maketitle
\flushbottom

\section{\texorpdfstring{Introduction}{Introduction}}
\label{sec:intro}

A foundational result in modern physics is that the vacuum state of a quantum field is an observer-dependent concept. Observers in non-inertial motion or in the presence of event horizons perceive the Minkowski vacuum as a thermal bath of particles, a phenomenon that underpins both the Unruh and Hawking effects \cite{Rindler66, Fulling1973, Unruh1976, Hawking1975, Davies1975}. This thermality can be understood from two complementary perspectives: one kinematic, the other dynamic.

From the kinematic viewpoint, thermality arises from the mathematical structure of the quantum state itself. By quantizing a field in a non-inertial coordinate system, such as Rindler coordinates for a uniformly accelerated observer, one finds that the inertial vacuum appears as a many-particle state with a precisely thermal distribution \cite{Unruh1976, Takagi1986}. This is encoded in the Bogoliubov transformation relating the different mode bases, and the thermal nature of the state is formally guaranteed by the KMS condition, a property rooted in the analytic structure of the vacuum’s correlation functions \cite{Martin_Schwinger1959, Kubo1957, Bisognano_Wichmann1975, Sewell1982}. This line of thought naturally leads to “generalized” or deformed vacuum states, defined purely by their characteristic mode mixing and correlations, often without reference to a specific spacetime geometry or dynamical mechanism that might produce them.

From the dynamic viewpoint, these thermal effects find a powerful and exactly solvable model in the form of an accelerating boundary, or moving mirror, in $(1+1)$ dimensions. Originating in the study of quantum fields in cavities with time-dependent boundaries \cite{Moore1970}, the moving mirror model became a crucial theoretical laboratory for understanding particle creation in curved spacetimes, serving as a simplified analogue for black hole evaporation \cite{Fulling_Davies1976, Fulling_Davies1977, Birrell_Davies1982}. In this framework, the full dynamical content of the particle creation process is encoded in the mirror’s trajectory, specified by a ray-tracing map $v=p(u)$. For a conformal field theory, the resulting asymptotic energy flux on future null infinity is elegantly determined by the Schwarzian derivative of this map, $\{p,u\}$. A particularly important trajectory is the exponential path investigated by Carlitz and Willey, which generates a constant, thermal energy flux at a temperature set by the mirror’s acceleration parameter $\kappa$, providing a steady analogue of Hawking radiation \cite{Carlitz_Willey1987}. This conceptually simple model remains a vibrant area of research, connecting quantum vacuum radiation to the dynamical Casimir effect and analogue gravity experiments \cite{Svidzinsky2018PRL, Chen_Mourou2017PRL, Good2020Wilczek}.

Motivated by the kinematic viewpoint, a significant body of recent work has developed families of generalized vacua parameterized by a continuous parameter $\kappa$, defined by their characteristic mode mixing. In Rindler spacetime, this led to the construction of a $\kappa$-Rindler vacuum built from linear combinations of Rindler modes, interpolating between the standard Rindler vacuum and the Minkowski vacuum in a controlled way and yielding modified Unruh physics for particle detectors \cite{Azizi2022Kappashort, Azizi2023JHEP, Azizi2025Tunable}. The construction was subsequently adapted to inertial frames, defining the \textit{$\kappa$-plane-wave vacuum} from linear combinations of Minkowski plane-wave modes \cite{Azizi2025KappaPW}. For each fixed $\kappa>0$ this procedure yields a \emph{distinct} vacuum state $\ket{0_\kappa}$, which is neither the Minkowski vacuum nor the Rindler vacuum for any finite value of $\kappa$. The only standard vacuum recovered within this family is the Minkowski vacuum in the singular limit $\kappa\to0$, where $\ket{0_\kappa}$ reduces to $\ket{0_{\rm M}}$. Further generalizations have incorporated non-stationary, phase-dependent particle creation \cite{Azizi2025KappaGamma}. In all these formulations, the $\kappa$-states are fundamentally kinematic constructions: they are defined by their structure, without being tied \emph{a priori} to a specific boundary condition or dynamical process that generates them.

In this paper, we bridge this gap by providing a concrete physical realization for the $\kappa$-plane-wave vacuum. We demonstrate that the quantum state of the right-moving radiation field produced by a Carlitz--Willey mirror with acceleration parameter $\kappa$ is physically and operationally identical to the $\kappa$-plane-wave vacuum on future right null infinity $\mathscr{I}^+_{R}$. Our analysis shows that the Carlitz--Willey mirror supplies an explicit dynamical mechanism that produces exactly the kinematic state previously defined in terms of mode mixing alone.

We establish this equivalence through a three-pronged proof:
\begin{enumerate}
    \item \emph{Wightman functions and KMS condition:} We show that the chiral two-point functions for both systems possess the same non-local thermal kernel, a universal $\ln\sinh$ structure that satisfies the KMS condition at inverse temperature $2\pi/\kappa$.

    \item \emph{Bogoliubov transformations:} We demonstrate that the Bogoliubov transformation for the Carlitz--Willey mirror, when diagonalized in a natural Mellin (log-frequency) basis, becomes a single-mode squeeze with the same Boltzmann factor $|\beta|^2/|\alpha|^2=e^{-2\pi\omega/\kappa}$ that characterizes the $\kappa$-plane-wave vacuum relative to the Minkowski state.

    \item \emph{Detector response:} We show that an inertial Unruh--DeWitt \cite{Unruh1976, Einstein100, Colosi2009Rovelli} detector coupled to the right-moving sector registers an identical Planckian thermal response at temperature $T=\kappa/(2\pi)$ in both backgrounds.
\end{enumerate}
Taken together, these results confirm that the Carlitz--Willey mirror provides the precise dynamical mechanism that generates the state described by the $\kappa$-plane-wave vacuum. This work grounds a formal kinematic construction in a well-understood physical model, clarifying the connection between kinematic and dynamic descriptions of quantum thermal phenomena.

The paper is organized as follows. Section~\ref{sec:bogol} details the Bogoliubov transformations for both systems and proves their equivalence. Section~\ref{sec:wightman} derives and compares the Wightman functions, while Section~\ref{sec:udw} calculates the detector responses. Section~\ref{sec:generalizations} generalizes our findings by deriving the complete class of mirror trajectories that produce the same asymptotic thermal state. We conclude in Section~\ref{sec:conclusion}.

\section{\texorpdfstring{Bogoliubov transformations}{Bogoliubov transformations}} \label{sec:bogol}

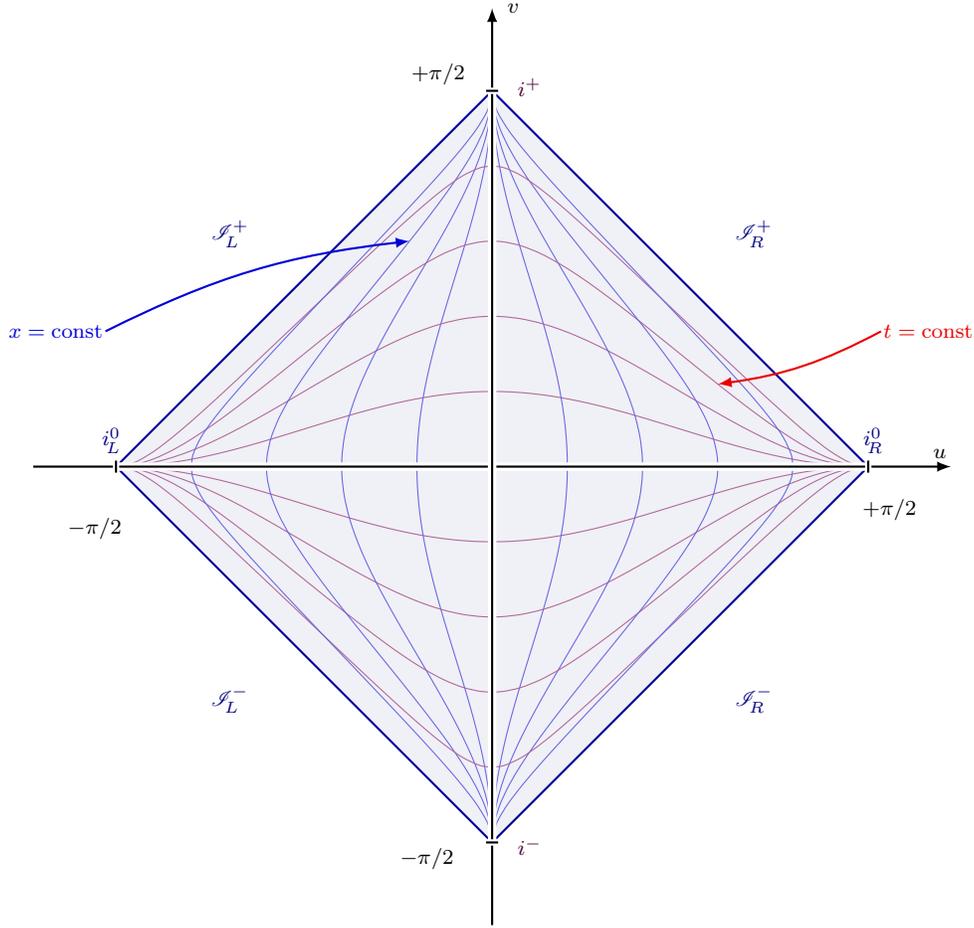
\begin{figure}[t!]
    \centering
\begin{tikzpicture}[scale=5, every node/.style={font=\scriptsize}]
  \message{Penrose diagram (cleaned, color-coded)^^J}

  \colorlet{constblue}{blue!85!black}
  \colorlet{constred}{red!92!black} 

  \tikzset{
    axisline/.style={thick, draw=black, preaction={draw=white, line width=3.2pt}},
    tickline/.style={thick, draw=black, preaction={draw=white, line width=2.4pt}},
  }

  \def\Nlines{4}
  \def\ta{tan(90*1.0/(\Nlines+1))}
  \def\tb{tan(90*2.0/(\Nlines+1))}
  \def\ctA{tan(90*3/(\Nlines+1))} 

  \coordinate (O) at ( 0, 0);
  \coordinate (S) at ( 0,-1);
  \coordinate (N) at ( 0, 1);
  \coordinate (W) at (-1, 0);
  \coordinate (E) at ( 1, 0);

  \coordinate (X)  at ({penroseu(\tb,\tb)},{penrosev(\tb,\tb)});
  \coordinate (X0) at ({penroseu(\ta,-\tb)},{penrosev(\ta,-\tb)});

  \fill[mylightblue] (N) -- (E) -- (S) -- (W) -- cycle;

  \draw[world line] (N) -- (S);
  \draw[world line] (W) -- (E);
  \foreach \i [evaluate={\c=\i/(\Nlines+1); \ct=tan(90*\c);}] in {1,...,\Nlines}{
    \draw[world line t,samples=\Nsamples,smooth,variable=\t,domain=-1:1]
      plot(\t,{-penrose(\t*pi/2,\ct)})
      plot(\t,{ penrose(\t*pi/2,\ct)});
    \draw[world line,samples=\Nsamples,smooth,variable=\x,domain=-1:1]
      plot({-penrose(\x*pi/2,\ct)},\x)
      plot({ penrose(\x*pi/2,\ct)},\x);
  }

  \draw[thick,blue!60!black] (N) -- (E) -- (S) -- (W) -- cycle;

  \node[mydarkblue,anchor=west, xshift=4, fill=white, inner sep=1pt, rounded corners=1pt]  at (.950, 0.07) {$i^0_{\!R}$};
  \node[mydarkblue,anchor=east, xshift=-4, fill=white, inner sep=1pt, rounded corners=1pt] at (-.950, 0.07) {$i^0_{\!L}$};

  \node[mydarkpurple,anchor=south, yshift=4, fill=white, inner sep=1pt, rounded corners=1pt] at (0.1, .95) {$i^+$};
  \node[mydarkpurple,anchor=north, yshift=-4, fill=white, inner sep=1pt, rounded corners=1pt] at (0.1, -.95) {$i^-$};

  \node[mydarkblue,anchor=west] at ( 0.62, 0.62) {$\mathscr I^+_{\!R}$};
  \node[mydarkblue,anchor=west] at ( 0.62,-0.62) {$\mathscr I^-_{\!R}$};
  \node[mydarkblue,anchor=east] at (-0.62, 0.62) {$\mathscr I^+_{\!L}$};
  \node[mydarkblue,anchor=east] at (-0.62,-0.62) {$\mathscr I^-_{\!L}$};

  \coordinate (tpt) at ( 0.60, { penrose(0.60*pi/2, \ctA) } );            
  \coordinate (xpt) at ({-penrose(0.60*pi/2, \ctA)}, 0.60);               

  \node[text=constred, fill=white, inner sep=0.8pt, rounded corners=1pt] (tlabel) at (1.16, 0.36) {$t=\text{const}$};
  \draw[->,>=latex,thick,constred] (tlabel.west) to[bend left=10] (tpt);

  \node[text=constblue, fill=white, inner sep=0.8pt, rounded corners=1pt] (xlabel) at (-1.16, 0.36) {$x=\text{const}$};
  \draw[->,>=latex,thick,constblue] (xlabel.east) to[bend left=10] (xpt);

  \draw[axisline,->] (-1.22,0) -- (1.22,0) node[below right=-.53] {$u$};
  \draw[axisline,->] (0,-1.22) -- (0,1.22) node[left=-.5] {$v$};

  \draw[tickline] ( 1,0) ++(0,0.016) -- ++(0,-0.032);
  \node[below right=-1] at ( 1.1, -0.2) {$+\pi/2$};

  \draw[tickline] (-1,0) ++(0,0.016) -- ++(0,-0.032);
  \node[below left=-1]  at (-1.1, -0.25) {$-\pi/2$};

  \draw[tickline] (0, 1) ++(0.016,0) -- ++(-0.032,0);
  \node[above right=-1] at (-0.10, 1.13) {$+\pi/2$};

  \draw[tickline] (0,-1) ++(0.016,0) -- ++(-0.032,0);
  \node[below right=-1] at (-0.13,-1.13) {$-\pi/2$};


\end{tikzpicture}

    \caption{\textbf{Penrose diagram of (1+1)D Minkowski spacetime.} The conformal diagram brings the asymptotic boundaries of spacetime to a finite location, illustrating the global causal structure. The Bogoliubov transformation relates the ``in'' quantum state, defined on past null infinity ($\mathscr{I}^-_{\!L} \cup \mathscr{I}^-_{\!R}$), to the ``out'' state measured on future null infinity ($\mathscr{I}^+_{\!L} \cup \mathscr{I}^+_{\!R}$). The diagram also shows lines of constant time ($t$) and constant position ($x$), which become hyperbolae in these conformal coordinates.}
    \label{fig:penrose-minkowski-bogol}
\end{figure}

The thermal phenomena at the heart of this paper originate from the mismatch between different notions of positive frequency, which is mathematically encoded by a Bogoliubov transformation. In this section, we derive the Bogoliubov transformations for both the kinematic $\kappa$-plane-wave vacuum and the dynamic Carlitz--Willey mirror. We show that while they arise from different physical starting points, they are fundamentally equivalent, each describing the same single-mode thermal squeeze.

\subsection{\texorpdfstring{The $\boldsymbol{\kappa}$-plane-wave vacuum: a single-mode squeeze}{The kappa-plane-wave vacuum: a single-mode squeeze}}

The relationship between any two such bases, parameterized by $\kappa$ and $\kappa'$, is a Bogoliubov transformation. Expanding the quantum field in each basis, we can match the coefficients of the underlying plane waves $e^{\mp i\Lambda u}$ to find that the transformation reduces to a simple, \emph{single-mode} SU(1,1) squeeze for each frequency $\Lambda>0$:
\begin{align}
{\cal A}_{\Lambda,\kappa'}
=&\alpha_\Lambda(\kappa',\kappa)\,{\cal A}_{\Lambda,\kappa}
+\beta_\Lambda(\kappa',\kappa)\,{\cal A}^{\dagger}_{\Lambda,\kappa},
\nn\\
\alpha_\Lambda=&\frac{\sinh(Y+X)}{\sqrt{\sinh(2X)\,\sinh(2Y)}},\quad
\beta_\Lambda=\frac{\sinh(Y-X)}{\sqrt{\sinh(2X)\,\sinh(2Y)}},
\label{eq:kappa-to-kappa}
\end{align}
with the shorthand $X:=\pi\Lambda/(2\kappa)$ and $Y:=\pi\Lambda/(2\kappa')$. These coefficients satisfy the canonical relation $|\alpha_\Lambda|^2-|\beta_\Lambda|^2=1$. The transformation can be parameterized by a squeeze parameter $r_\Lambda$, where
\begin{align}
\tanh r_\Lambda=&\frac{\beta_\Lambda}{\alpha_\Lambda}
=\frac{\sinh(Y-X)}{\sinh(Y+X)}
=\frac{\tanh Y-\tanh X}{\tanh Y+\tanh X},
\end{align}
The most important case is the transformation between the $\kappa$-basis and the standard Minkowski basis, which corresponds to the limit $\kappa' \to 0$ (or $Y \to \infty$). In this limit, the ratio of the Bogoliubov coefficients yields the characteristic thermal Boltzmann weight:
\begin{align}
\frac{|\beta_\Lambda|^2}{|\alpha_\Lambda|^2}
=\frac{\sinh^2(Y-X)}{\sinh^2(Y+X)}\xrightarrow{\,Y\to\infty\,}e^{-4X}
=e^{-2\pi\Lambda/\kappa}.
\end{align}
This implies that the $\kappa$-vacuum state $|0_\kappa\rangle$, defined by ${\cal A}_{\Lambda,\kappa}|0_\kappa\rangle = 0$, contains a thermal distribution of Minkowski particles. The mean number of Minkowski quanta in the $\kappa$-vacuum is given by
\begin{align}
\langle 0_\kappa| a_\Lambda^\dagger a_\Lambda |0_\kappa \rangle = |\beta_\Lambda(\kappa, 0)|^2 = \frac{1}{e^{2\pi\Lambda/\kappa}-1},
\end{align}
which is a perfect Planckian spectrum at the temperature $T_\kappa = \kappa/(2\pi)$.

\subsection{\texorpdfstring{The Carlitz--Willey mirror: Mellin diagonalization}{The Carlitz-Willey mirror: Mellin diagonalization}}

\begin{figure}[t]
    \centering
    \includegraphics[width=.8\textwidth]{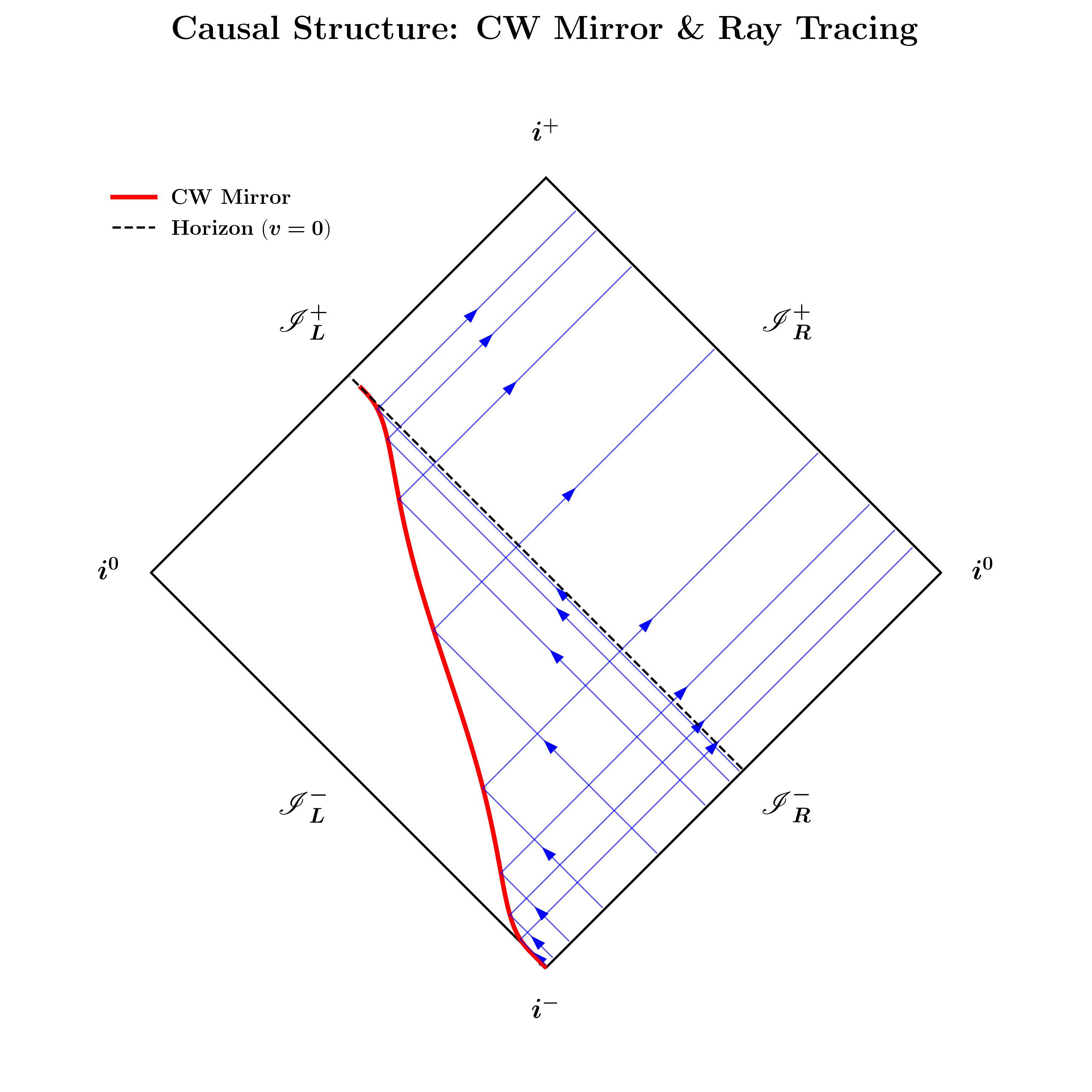}
    \caption{Causal structure and ray tracing for the Carlitz--Willey 
    trajectory on a compactified Minkowski diagram. 
    The curved worldline represents the mirror, which emerges from past 
    timelike infinity $i^{-}$ and asymptotically approaches the null line 
    $v=v_{H}$ (dashed), defining a future event horizon. 
    Incoming right–moving null rays from $\mathscr I^-_{\!R}$ reflect from 
    the mirror and are bunched toward $\mathscr I^+_{\!R}$, producing a 
    steady thermal flux; rays that cross the horizon never reach 
    $\mathscr I^+_{\!R}$ and instead end on $\mathscr I^+_{\!L}$.}
    \label{fig:cw-penrose}
\end{figure}

For the Carlitz--Willey mirror, the Bogoliubov transformation relates the ``in'' modes, defined on past null infinity, to the ``out'' modes, defined on future null infinity. The trajectory is specified by the exponential ray-tracing map
\begin{align}
v=p(u)=v_H - A\,e^{-\kappa u},
\qquad A>0,\;\kappa>0,
\label{eq:CW-raytracing}
\end{align}
so that the mirror worldline emerges from $i^{-}$ with $u\to-\infty,\ v\to-\infty$ and asymptotically approaches the null horizon $v=v_H$ as $u\to+\infty$ (see Fig.~\ref{fig:cw-penrose}). The corresponding worldline $x=z(t)$ and its kinematics are derived in detail in App.~\ref{app:tx-mirror}. In App.~\ref{app:inner-product} we construct the properly normalized ``in'' and ``out'' modes adapted to \eqref{eq:CW-raytracing}, compute their Klein--Gordon inner products on null Cauchy surfaces, and obtain the general integral expressions for the Bogoliubov kernels.

For the CW trajectory it is convenient to shift $v$ so that the horizon lies at $v_H=0$ and work with
\begin{align}
p(u)=-A\,e^{-\kappa u},
\qquad A=\frac{1}{\kappa}>0,
\end{align}
as in App.~\ref{app:CW-kernels}. In that case the Bogoliubov kernels
$\alpha_{\omega\omega'}$ and $\beta_{\omega\omega'}$ simplify to the
factorized form
\begin{align}
\alpha_{\omega\omega'}
&=F(\omega)\,\omega'^{-1/2+i\omega/\kappa},\qquad
\beta_{\omega\omega'}
=G(\omega)\,\omega'^{-1/2+i\omega/\kappa},
\label{eq:alpha-beta-factorized-body}
\end{align}
with
\begin{align}
F(\omega)
&=\frac{1}{2\pi}\sqrt{\frac{1}{\omega}}\,
A^{i\omega/\kappa}\,e^{+\pi\omega/(2\kappa)}\,
\Gamma\!\left(1-\frac{i\omega}{\kappa}\right),\\[1mm]
G(\omega)
&=\frac{1}{2\pi}\sqrt{\frac{1}{\omega}}\,
A^{i\omega/\kappa}\,e^{-\pi\omega/(2\kappa)}\,
\Gamma\!\left(1-\frac{i\omega}{\kappa}\right),
\label{eq:F-G-def-body}
\end{align}
as derived in App.~\ref{app:CW-kernels}. All of the $\omega'$--dependence is carried by the Mellin factor $\omega'^{-1/2+i\omega/\kappa}$.

This suggests introducing a Mellin (log-frequency) basis for the \emph{in}–operators,
\begin{align}
b^{\rm in}_{\Omega}
=\int_{0}^{\infty}\frac{d\omega'}{\sqrt{2\pi\kappa}}\,
\omega'^{-\frac12 - i\Omega/\kappa}\,a^{\rm in}_{\omega'},
\qquad \Omega\in\mathbb{R},
\label{eq:Mellin-b-body}
\end{align}
with inverse
\begin{align}
a^{\rm in}_{\omega'}
=\int_{-\infty}^{+\infty}\frac{d\Omega}{\sqrt{2\pi\kappa}}\,
\omega'^{-\frac12 + i\Omega/\kappa}\,b^{\rm in}_{\Omega}.
\end{align}
The Mellin orthogonality relation,
\begin{align}
\int_{0}^{\infty}\frac{d\omega'}{2\pi\kappa}\,
\omega'^{-1+i(\Omega-\Omega')/\kappa}
=\delta(\Omega-\Omega'),
\label{eq:Mellin-orthogonality-body}
\end{align}
implies
\begin{align}
[b^{\rm in}_{\Omega},b^{\rm in\,\dagger}_{\Omega'}]
=\delta(\Omega-\Omega'),
\end{align}
so the $b^{\rm in}_{\Omega}$ form a canonical bosonic basis (see App.~\ref{app:Mellin-Bogol} for details).

Substituting the inverse Mellin transform into the usual Bogoliubov relation
\begin{align}
a^{\rm out}_{\omega}
=\int_{0}^{\infty}\!d\omega'\,
\Big(\alpha_{\omega\omega'}\,a^{\rm in}_{\omega'}
+\beta_{\omega\omega'}\,a^{\rm in\,\dagger}_{\omega'}\Big),
\label{eq:usual-Bogol-body}
\end{align}
and using the factorization
\eqref{eq:alpha-beta-factorized-body} together with
\eqref{eq:Mellin-orthogonality-body}, the $\omega'$–integrals collapse.
As worked out explicitly in App.~\ref{app:Mellin-Bogol}, one finds
\begin{align}
a^{\rm out}_{\omega}
=F(\omega)\,b^{\rm in}_{\omega}
+G(\omega)\,b^{\rm in\,\dagger}_{-\omega},
\label{eq:aout-in-Mellin-body}
\end{align}
with no residual integral over frequencies. Each real label $\omega$ therefore defines an independent SU(1,1) block acting on the two-dimensional subspace spanned by the Mellin pair
\begin{align}
\big(b^{\rm in}_{\omega},\,b^{\rm in\,\dagger}_{-\omega}\big).
\end{align}
In this precise sense the CW Bogoliubov transformation is ``single–mode'' in the Mellin basis: every out–mode $a^{\rm out}_{\omega}$ couples only to the pair $(\omega,-\omega)$ and not to any other Mellin labels.

From \eqref{eq:F-G-def-body} it follows immediately that
\begin{align}
\frac{|G(\omega)|^{2}}{|F(\omega)|^{2}}
=e^{-2\pi\omega/\kappa},
\label{eq:CW-thermal-ratio-body}
\end{align}
so, after a conventional frequency–dependent normalization of the out–basis (absorbed into a redefinition of $a^{\rm out}_{\omega}$), the SU(1,1) coefficients in \eqref{eq:aout-in-Mellin-body} satisfy the usual relation $|\alpha_{\omega}|^{2}-|\beta_{\omega}|^{2}=1$ with
\begin{align}
\frac{|\beta_{\omega}|^{2}}{|\alpha_{\omega}|^{2}}
=e^{-2\pi\omega/\kappa}.
\end{align}
Since the ``in'' vacuum is annihilated by $a^{\rm in}_{\omega'}$ and hence by all $b^{\rm in}_{\Omega}$, the mean number of outgoing particles of frequency $\omega$ created by the mirror is
\begin{align}
\langle 0_{\rm in}|\,a^{\rm out\,\dagger}_{\omega}a^{\rm out}_{\omega}\,|0_{\rm in}\rangle
=\frac{1}{e^{2\pi\omega/\kappa}-1},
\label{eq:CW-Nomega-body}
\end{align}
showing that the CW mirror radiates a perfect thermal spectrum at temperature
\begin{align}
T_{\rm CW}=\frac{\kappa}{2\pi}.
\end{align}

Hence we find that the CW mirror implements, in the Mellin basis, a family
of independent SU(1,1) Bogoliubov transformations labelled by $\omega$, with
thermal ratio $|\beta_\omega|^{2}/|\alpha_\omega|^{2}=e^{-2\pi\omega/\kappa}$
and Planckian occupation number 
$\langle 0_{\rm in}|a^{\rm out\,\dagger}_\omega a^{\rm out}_\omega|0_{\rm in}\rangle
=1/(e^{2\pi\omega/\kappa}-1)$. Comparing with the $\kappa$-plane-wave map
to Minkowski, which yields the same SU(1,1) structure with
$|\beta_\Lambda|^{2}/|\alpha_\Lambda|^{2}=e^{-2\pi\Lambda/\kappa}$ and
$\langle 0_\kappa|a^{\dagger}_\Lambda a_\Lambda|0_\kappa\rangle
=1/(e^{2\pi\Lambda/\kappa}-1)$, we see that, relative to the Minkowski
vacuum, the kinematic $\kappa$-vacuum and the dynamic CW mirror are
Bogoliubov–equivalent realizations of the same thermal single–mode squeeze
at temperature $T=\kappa/(2\pi)$.

\section{\texorpdfstring{Wightman functions}{Wightman functions}}
\label{sec:wightman}

Having established the equivalence at the level of the operator algebra, we now demonstrate it at the level of the two-point correlation functions. We derive the Wightman functions for both the $\kappa$-plane-wave vacuum and the Carlitz-Willey mirror state, showing that their physically significant nonlocal parts, which dictate the thermal properties on future null infinity, are identical.

\subsection{\texorpdfstring{The $\boldsymbol{\kappa}$-plane-wave vacuum}{The kappa-plane-wave vacuum}}
\label{sec:wightman-kappa}

The Wightman function for the right-moving field in the $\kappa$-plane-wave vacuum, defined by $\mathcal{A}_\Lambda|0_\kappa\rangle = 0$, is given by $W^{>}_\kappa(u,u')=\bra{0_\kappa}\Phi_{\rm RTW}(u)\Phi_{\rm RTW}(u')\ket{0_\kappa}$. The full calculation (see App.~\ref{app:Wightman_kappa}) shows that the two-point function decomposes into a stationary piece dependent only on the separation $\Delta u:=u-u'$, and a non–stationary, local piece dependent on the sum $u+u'$. The exact result is
\begin{align}
W^{\rm RTW}_\kappa(u,u')
&=-\frac{1}{4\pi}\,\ln\!\Big[\mu^2\,\frac{2}{\kappa}\,
\sinh\!\Big(\frac{\kappa}{2}\,(\Delta u-i\epsilon)\Big)\Big]
\;-\;\frac{1}{4\pi}\,\ln\!\cosh\!\Big(\frac{\kappa}{2}\,(u+u')\Big), \label{eq:Wkappa-RTW-exact}
\end{align}
and $W^{<}_\kappa$ is obtained by the replacement $i\epsilon \to -i\epsilon$ in the stationary term. For the purposes of analyzing thermality, one isolates the stationary part,
\begin{align}
W^{>}_{\kappa, \text{stationary}}(\Delta u)
&= -\frac{1}{4\pi}\,
\ln\!\Big[\mu^2\,\frac{2}{\kappa}\,
\sinh\!\Big(\frac{\kappa}{2}\,(\Delta u-i\epsilon)\Big)\Big],  
\label{eq:Wkappa-greater}\\[2mm]
W^{<}_{\kappa, \text{stationary}}(\Delta u)
&= -\frac{1}{4\pi}\,
\ln\!\Big[\mu^2\,\frac{2}{\kappa}\,
\sinh\!\Big(\frac{\kappa}{2}\,(\Delta u+i\epsilon)\Big)\Big],
\label{eq:Wkappa-less}
\end{align}
which manifestly obeys the KMS condition at temperature $T=\kappa/2\pi$:
\begin{align}
W^{>}_{\kappa, \text{stationary}}(\Delta u-i\,2\pi/\kappa)=W^{<}_{\kappa, \text{stationary}}(\Delta u).
\end{align}
In the $\kappa\to0$ limit, the stationary part correctly reduces to the Minkowski correlator, $-(4\pi)^{-1}\ln[\mu^2(\Delta u \mp i\epsilon)]$, while the non-stationary part vanishes as $\mathcal{O}(\kappa^2)$.

\begin{figure}[ht!]
    \centering
    \includegraphics[width=\textwidth]{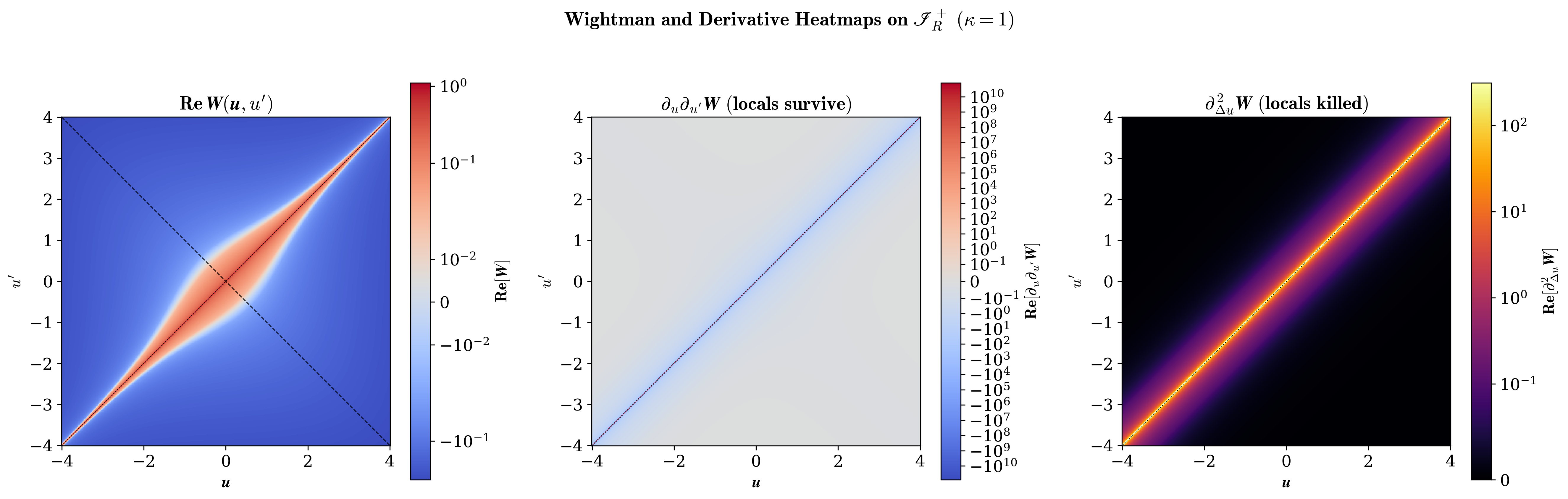}
    \caption{\textbf{Structure of the Wightman function on $\mathscr{I}^+_R$ for the $\kappa$-plane-wave vacuum ($\kappa=1$).} All plots use a symmetric logarithmic color scale to handle the singularity on the diagonal and reveal off-diagonal features.
    \textbf{(Left)} The full Wightman function $\mathrm{Re}\,W(u,u')$. Its structure shows dependence on both the non-local separation $\Delta u = u-u'$ (features parallel to the main diagonal) and the local coordinate sum $\Sigma = u+u'$ (features perpendicular to the diagonal).
    \textbf{(Center)} The mixed derivative $\partial_u \partial_{u'} W$. In this plot, the local, non-stationary part survives, visible as the cross-like structure that depends on $u+u'$.
    \textbf{(Right)} The universal thermal kernel $\partial_{\Delta u}^2 W$. This derivative operator eliminates the local, $\Sigma$-dependent part, leaving a pure function of $\Delta u$ that is constant along lines perpendicular to the diagonal. This visually confirms the isolation of the stationary, non-local thermal correlations.}
    \label{fig:wightman_heatmaps}
\end{figure}

\subsection{\texorpdfstring{The Carlitz--Willey mirror}{The Carlitz-Willey mirror}}
\label{sec:wightman-cw}

\begin{figure}[t!]
    \centering
    \includegraphics[width=.8\textwidth]{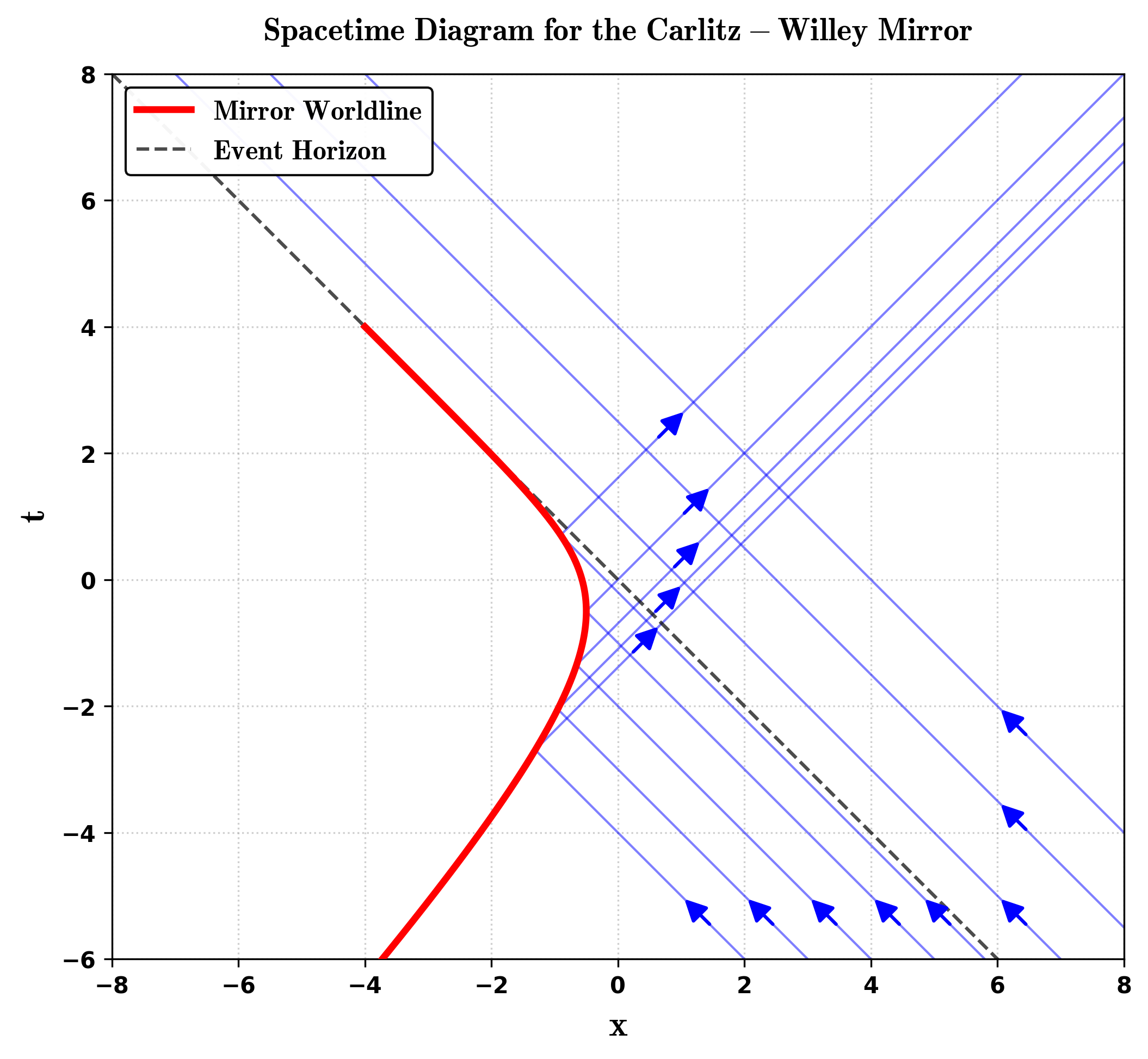}
    \caption{\textbf{Minkowski spacetime diagram of the accelerating Carlitz-Willey mirror.} The mirror's worldline (red curve) undergoes eternal acceleration, asymptotically approaching the speed of light and creating an event horizon (dashed line) behind it. Uniformly spaced incoming null rays (blue lines from the top-left), representing the empty Minkowski vacuum, reflect off the mirror. Due to the mirror's increasing velocity, the outgoing reflected rays are compressed, or "bunched up." An asymptotic observer on $\mathscr{I}^+_R$ (at the far right) intercepts this compressed wave pattern, which has the power spectrum of thermal radiation at a temperature $T=\kappa/(2\pi)$ set by the mirror's acceleration.}
    \label{fig:cw_minkowski}
\end{figure}

For the mirror, the Wightman function $W^{>}(u,v;u',v')=\bra{0_{\rm in}}\Phi(u,v)\Phi(u',v')\ket{0_{\rm in}}$ is constructed by summing over the ``in''-modes. This yields the standard image form for the bulk two-point function:
\begin{empheq}[box=\widefbox]{align}
W^{>}(u,v;u',v')
=& -\frac{1}{4\pi}\,
\ln\!\left[
\frac{(v-v'-i\epsilon)\,\big(p(u)-p(u')-i\epsilon\big)}
{\big(v-p(u')-i\epsilon\big)\,\big(p(u)-v'-i\epsilon\big)}
\right].
\label{eq:CW-image-form}
\end{empheq}
To compare with the chiral $\kappa$-vacuum, we must take the chiral limit on future right null infinity ($\mathscr I^+_R$). By holding $u,u'$ fixed and sending $v,v'\to+\infty$ with $v-v'$ finite, the mixed terms involving both $u$ and $v$ cancel (up to local terms absorbed into the scale $\mu$), leaving a purely right-moving correlator whose non-local structure is determined by $p(u)-p(u')$. For the specific Carlitz-Willey trajectory, $p(u)=-A e^{-\kappa u}$, we have
\begin{align}
p(u)-p(u')
=-2A\,e^{-\frac{\kappa}{2}(u+u')}\,
\sinh\!\Big(\tfrac{\kappa}{2}(u-u')\Big).
\end{align}
The resulting chiral Wightman function on $\mathscr I^+_R$ is therefore
\begin{empheq}[box=\widefbox]{align}
W^{(\mathscr I^+_R)\,>}(u,u')
= -\frac{1}{4\pi}\,
\ln\!\Bigg[
\mu^2\,\frac{p(u)-p(u')-i\epsilon}{\,u-u'-i\epsilon\,}
\Bigg].
\label{eq:equiv-Wcw-chiral}
\end{empheq}
After substituting the expression for $p(u)-p(u')$, the $e^{-\frac{\kappa}{2}(u+u')}$ factor contributes only a local term, while the $\sinh(\frac{\kappa}{2}(u-u'))$ term provides the non-local thermal dependence.

\subsection{\texorpdfstring{Equivalence on $\boldsymbol{\mathscr I^+_R}$ and Bulk Inequivalence}{Equivalence on I+R and Bulk Inequivalence}}

Comparing the results from the two systems reveals their precise relationship. The stationary part of the $\kappa$-plane-wave Wightman function, Eq.~\eqref{eq:Wkappa-greater}, is
\begin{align}
W^{>}_{\kappa, \text{stationary}}(\Delta u) = -\frac{1}{4\pi} \ln\!\Bigg[\mu^2\,\frac{2}{\kappa}\, \sinh\!\Big(\frac{\kappa}{2}(\Delta u-i\epsilon)\Big)\Bigg].
\end{align}
For the CW mirror, the chiral limit of its Wightman function, after accounting for the local $u+u'$ terms, has the exact same non-local dependence:
\begin{align}
W^{(\mathscr I^+_R)\,>}(u,u') = -\frac{1}{4\pi}\, \ln\!\Bigg[ \mu^2\,\frac{2}{\kappa}\, \sinh\!\Big(\tfrac{\kappa}{2}(u-u'-i\epsilon)\Big) \Bigg] +\text{local}.
\end{align}
Thus, both descriptions share exactly the same KMS-thermal kernel on $\mathscr I^+_R$. A direct physical consequence of this equivalence is that they produce the same renormalized energy flux. Point-splitting the Wightman function, or equivalently using the Schwarzian derivative of the mirror map, gives the constant thermal plateau:
\begin{empheq}[box=\widefbox]{align}
F(u)=\langle T_{uu}\rangle
= -\frac{1}{24\pi}\{p,u\}
= \frac{\kappa^{2}}{48\pi}.
\end{empheq}
However, it is crucial to recognize that this equivalence is sectoral and asymptotic. For generic bulk points, the two states are different. The full CW Wightman function, Eq.~\eqref{eq:CW-image-form}, is not a sum of independent left- and right-moving functions due to the cross terms like $\ln(v-p(u'))$. Consequently, it is not stationary and does not satisfy the KMS condition with respect to inertial time in the bulk, a key distinction from the factorized $\kappa$-plane-wave vacuum state.

\section{\texorpdfstring{Unruh--DeWitt detector responses}{Unruh-DeWitt detector responses}}
\label{sec:udw}

\noindent\emph{Scope.} The final proof of equivalence comes from a direct physical observable: the excitation rate of a particle detector. We consider an inertial Unruh--DeWitt  detector with a derivative coupling to probe the right–moving sector on $\mathscr I^+_R$. In the long-time (adiabatic) limit, we show that the stationary excitation rate is identically Planckian in both the $\kappa$–plane–wave vacuum and the Carlitz--Willey (CW) background. This agreement pertains to the \textbf{right–moving, asymptotic response} and does not assert equality of bulk correlators or KMS with respect to inertial time in the interior.

\subsection{\texorpdfstring{Response in the $\boldsymbol{\kappa}$-plane-wave vacuum}{Response in the kappa-plane-wave vacuum}}

We first consider a two–level UDW detector with energy gap $\omega>0$ on an inertial worldline at fixed $x=x_0$, coupled to the right-moving field via the interaction Hamiltonian
\begin{align}
H_{\rm int}(t)=g\,\chi(t)\,m(t)\,\partial_t\Phi_{\rm RTW}(u)\Big|_{u=t-x_0},
\end{align}
where $m(t)$ is the detector's monopole moment and $\chi(t)$ is a long, smooth switching function. The detector's response can be calculated in two equivalent ways.

First, via first–order perturbation theory, the transition amplitude to a final state containing one $\kappa$-quantum of frequency $\Lambda$ is
\begin{align}
\mathcal A_{\Lambda}
=-\frac{i g}{\hbar}\int\!dt\,\chi(t)e^{+i\omega t}\,\partial_t\Phi^{*(\kappa)}_\Lambda(t-x_0),
\end{align}
which in the adiabatic limit $\widetilde{\chi}(\nu)\to2\pi\,\delta(\nu)$ becomes non-zero only for on-shell quanta with $\Lambda=\omega$. Squaring the amplitude and summing over all modes yields the Fermi–golden–rule rate:
\begin{align}
\mathcal R_{g\to e}
=\lim_{T\to\infty}\frac{P_{g\to e}}{T}
=\frac{g^2\omega}{2\hbar^2}\;
\frac{1}{e^{\frac{2\pi\omega}{\kappa}}-1}
\;\propto\;\frac{1}{e^{\omega/T_\kappa}-1},
\qquad T_\kappa=\frac{\kappa}{2\pi}.
\label{eq:kappa-rate-proj}
\end{align}
Alternatively, the response rate is given by the Fourier transform of the Wightman function, a quantity known as the response function $\dot{\mathcal F}(\omega)$. Using the stationary part of the $\kappa$-vacuum Wightman function,
\begin{align}
W^{>}_\kappa(s)=-\frac{1}{4\pi}\ln\!\Big[\mu^2\,\frac{2}{\kappa}\sinh\!\Big(\frac{\kappa}{2}(s-i\epsilon)\Big)\Big],
\qquad s:=t-t',
\end{align}
one finds the rate by integrating its second derivative:
\begin{align}
\dot{\mathcal F}_{\kappa}(\omega)
=\int_{-\infty}^{+\infty}\!ds\,e^{-i\omega s}\,\partial_s^2 W^{>}_\kappa(s)
=\frac{\omega}{2}\,\frac{1}{e^{\omega/T_\kappa}-1},
\qquad T_\kappa=\frac{\kappa}{2\pi}.
\label{eq:kappa-rate-w}
\end{align}
The Fourier transform of the thermal $\mathrm{csch}^2$ kernel directly yields the Bose factor. Both methods give the same Planckian response, and detailed balance follows immediately from the KMS property of the Wightman function.

\subsection{\texorpdfstring{Response in the Carlitz--Willey background}{Response in the Carlitz-Willey background}}

We now place the same inertial detector in the spacetime of a Carlitz--Willey mirror. As derived in Sec.~\ref{sec:bogol}, the ``in''-vacuum (Minkowski) evolves to an ``out''-state containing a thermal distribution of particles with mean occupation number
\begin{align}
\langle 0_{\rm in}|\,a^{\rm out\,\dagger}_{\Omega}a^{\rm out}_{\Omega}\,|0_{\rm in}\rangle
=N_\Omega=|\beta_{\Omega}|^{2}
=\frac{1}{e^{2\pi\Omega/\kappa}-1}.
\label{eq:CW-Nomega-recall}
\end{align}
The detector's excitation rate can be computed by finding the probability of absorbing one of these ``out''-quanta. The transition amplitude at first order is
\begin{align}
\mathcal A^{\rm(abs)}_{\Omega}
=-\frac{i g}{\hbar}\int_{-\infty}^{+\infty}\!dt\;\chi(t)\,e^{+i\omega t}\,
\partial_t\phi^{\rm out}_{\Omega}(t-x_0,t+x_0)\;
\langle N_{\Omega}\!-\!1|\,a^{\rm out}_{\Omega}\,|N_{\Omega}\rangle.
\end{align}
In the adiabatic limit, the time integral projects onto the on-shell component, and the total transition probability for excitation is
\begin{align}
P_{g\to e}
=\frac{g^2}{\hbar^2}\int_{0}^{\infty}\!d\Omega\;
\frac{\Omega}{4\pi}\,|\widetilde{\chi}(\omega-\Omega)|^{2}\,N_{\Omega}
\;\xrightarrow[]{\ T\to\infty\ }\;
\frac{g^2}{\hbar^2}\,\frac{\omega}{2}\,T\,N_{\omega},
\end{align}
where we have used $|\widetilde{\chi}|^{2}\to 2\pi T\,\delta(\omega-\Omega)$. The stationary rate is therefore directly proportional to the particle number $N_\omega$:
\begin{align}
\dot{\mathcal F}_{\rm CW}(\omega)
=\lim_{T\to\infty}\frac{P_{g\to e}}{T}
=\frac{g^2}{\hbar^2}\,\frac{\omega}{2}\,N_{\omega}.
\end{align}
Inserting the CW occupation number from Eq.~\eqref{eq:CW-Nomega-recall} yields the final Planckian response:
\begin{align}
\dot{\mathcal F}_{\rm CW}(\omega)
=\frac{g^2}{\hbar^2}\,\frac{\omega}{2}\,
\frac{1}{e^{2\pi\omega/\kappa}-1}
\propto\frac{\omega}{e^{\omega/T_{\rm CW}}-1},
\qquad T_{\rm CW}=\frac{\kappa}{2\pi}.
\end{align}
This result, derived from the particle-counting perspective, is identical to the one that would be obtained by using the chiral limit of the CW Wightman function, whose second derivative provides the same thermal $\mathrm{csch}^2$ kernel as in the $\kappa$-vacuum case.

\subsection{\texorpdfstring{Equivalence of Detector Response and the Minkowski Limit}{Equivalence of Detector Response and the Minkowski Limit}}

The calculations in the preceding subsections demonstrate a perfect agreement in the physically measurable response of an inertial detector. Both the kinematic $\kappa$-vacuum and the dynamic CW mirror cause the detector to thermalize at the exact same temperature $T=\kappa/(2\pi)$.
\begin{empheq}[box=\widefbox]{align}
\dot{\mathcal F}_{\kappa}(\omega) \propto \frac{\omega}{e^{\omega/(\kappa/2\pi)}-1} \quad \equiv \quad \dot{\mathcal F}_{\rm CW}(\omega).
\end{empheq}
This establishes the equivalence of the two systems in a third, independent channel, complementing the proofs from the Bogoliubov and Wightman perspectives.

This entire framework has a smooth and well-behaved Minkowski limit as $\kappa \to 0$. In this limit, the temperature $T \to 0$, and the detector response vanishes for any positive energy gap $\omega > 0$:
\begin{align}
\lim_{\kappa\to0} \dot{\mathcal F}(\omega) = 0.
\end{align}
The detector cools to its ground state, as expected in the empty Minkowski vacuum. This limit is consistent across all channels: the Bogoliubov coefficients $\beta \to 0$, the Wightman function becomes the standard Minkowski correlator, and the mirror's trajectory becomes trivial (inertial) after fixing an appropriate affine gauge, leading to zero particle flux. Hence, both the $\kappa$-plane-wave vacuum and the CW mirror state smoothly converge to the same, unique Minkowski endpoint.

\section{\texorpdfstring{Generalizations and Conditions for Thermality}{Generalizations and Conditions for Thermality}}
\label{sec:generalizations}

The equivalence established in the previous sections between the single-exponential Carlitz-Willey mirror and the $\kappa$-plane-wave vacuum raises a natural question: which class of mirror trajectories $v=p(u)$ reproduces this specific thermal physics? In this final section, we derive the complete family of trajectories that yield the universal thermal kernel on future null infinity and clarify the stricter conditions required to obtain a constant, stationary thermal flux.

For a general mirror, the right-moving Wightman function on $\mathscr I^+_{R}$ is
\begin{align}
W^{(\mathscr I^+_R)\,>}(u,u')
= -\frac{1}{4\pi}\ln\!\Bigg[\mu^2\,\frac{p(u)-p(u')-i\epsilon}{u-u'-i\epsilon}\Bigg].
\label{eq:chiral-general}
\end{align}
We demand that its nonlocal dependence on $\Delta u:=u-u'$ matches the $\kappa$-plane-wave kernel up to local, $u+u'$-dependent terms:
\begin{align}
p(u)-p(u') \;=\; H\!\Big(\tfrac{u+u'}{2}\Big)\;\sinh\!\Big(\tfrac{\kappa}{2}(u-u')\Big),
\qquad \forall\,u,u',
\label{eq:functional}
\end{align}
for some arbitrary function $H$. Expanding this functional relation in powers of $d=(u-u')/2$ reveals that it is equivalent to the linear ordinary differential equation
\begin{align}
p^{(3)}(u)=\kappa^2\,p'(u).
\label{eq:ode}
\end{align}
The general solution is a linear combination of exponentials:
\begin{align}
p(u)=\alpha e^{\kappa u}+\beta e^{-\kappa u}+\gamma,
\end{align}
with real constants $(\alpha,\beta,\gamma)$. Furthermore, the nonlocal structure of the Wightman function is invariant under Möbius reparametrizations of the coordinate $v$, $\tilde p(u)=(a\,p(u)+b)/(c\,p(u)+d)$. Thus, the complete class of trajectories that exactly reproduce the thermal kernel is given by the projective family
\begin{empheq}[box=\widefbox]{align}
\tilde p(u)=\frac{a(\alpha e^{\kappa u}+\beta e^{-\kappa u}+\gamma)+b}{c(\alpha e^{\kappa u}+\beta e^{-\kappa u}+\gamma)+d}\;,
\qquad ad-bc\neq0.
\label{eq:general-class}
\end{empheq}
This result can be further relaxed. If we only require the thermal kernel to emerge asymptotically for late times ($u,u'\to+\infty$), it is sufficient for the trajectory to have an exponential tail:
\begin{align}
p(u)=v_H - A\,e^{-\kappa u}+o(e^{-\kappa u})\qquad(u\to+\infty),\quad A>0.
\label{eq:asymp}
\end{align}
Finally, we consider the condition for a constant energy flux, $F(u)=-\tfrac{1}{24\pi}\{p,u\}$. The Schwarzian derivative for the general solution is
\begin{align}
\{p,u\}=\kappa^2-\frac{3}{2}\kappa^2\left(\frac{\alpha e^{\kappa u}+\beta e^{-\kappa u}}{\alpha e^{\kappa u}-\beta e^{-\kappa u}}\right)^{\!2},
\end{align}
which is constant (and equals $-\kappa^2/2$) only if one of the exponential terms is absent ($\alpha=0$ or $\beta=0$). This singles out the pure, single-exponential subclasses as the unique trajectories that produce a constant thermal plateau $F(u)=\kappa^2/(48\pi)$:
\begin{align}
p(u)=v_H - A\,e^{-\kappa u}\quad\text{(Carlitz–Willey type)},\qquad
p(u)=\tilde v_0 + \tilde A\,e^{+\kappa u}\quad\text{(time–reversed CW)}.
\end{align}
In summary, while a broad class of trajectories can generate an asymptotically thermal state, the condition of a perfectly stationary, constant thermal emission is much stricter, requiring a purely exponential ray-tracing map.

\section{\texorpdfstring{Conclusion}{Conclusion}}
\label{sec:conclusion}

In this paper, we have established a clear physical realization for the kinematic framework of the $\kappa$-plane-wave vacuum. We have demonstrated that this mathematical construction, previously defined by its characteristic thermal mode mixing, is not merely a formal convenience but is physically and operationally equivalent, in the right-moving chiral sector on future null infinity, to the quantum state produced by an accelerating Carlitz--Willey mirror. This result provides a concrete, dynamical origin for the thermal properties of the $\kappa$-vacuum.

Our proof of this equivalence is built on a three-pronged analysis, showing a precise match between the two systems in three independent and complementary channels. First, at the level of the operator algebra, we showed that the Bogoliubov transformation for the Carlitz--Willey mirror, when diagonalized in the natural Mellin (log-frequency) basis, reduces to a single-mode squeeze with the exact same thermal ratio $|\beta|^2/|\alpha|^2 = e^{-2\pi\omega/\kappa}$ that defines the $\kappa$-vacuum's relationship to the Minkowski vacuum. Second, we demonstrated that the chiral Wightman functions for both systems possess identical non-local thermal kernels on $\mathscr{I}^+_R$, both satisfying the KMS condition at a temperature $T=\kappa/(2\pi)$. This equivalence directly implies an identical, constant renormalized chiral energy flux on $\mathscr{I}^+_R$ for both systems. Finally, we showed that these foundational equivalences manifest in the same physically measurable observable: an inertial Unruh--DeWitt detector registers an identical, perfectly Planckian thermal response in both backgrounds in the asymptotic right-moving sector.

The significance of this result is that it provides a crucial bridge between the two canonical approaches to quantum thermal phenomena. It grounds the abstract, kinematic ``Route 1'' approach (quantization in non-inertial frames) in the concrete, dynamic ``Route 2'' approach (boundary-induced radiation). The parameter $\kappa$, once a purely mathematical label for a family of squeezed states, is now identified with a physical quantity: the acceleration or effective surface gravity that governs the mirror's exponential trajectory. Furthermore, our analysis in the final section generalized this result, deriving the complete class of mirror trajectories that generate the same universal $\kappa$-thermal kernel (exactly or asymptotically) on $\mathscr{I}^+_R$, and showing that the additional physical requirement of a stationary, constant flux uniquely singles out the single-exponential Carlitz--Willey trajectory.

This work opens several promising avenues for future investigation. The most immediate is to seek a dynamical realization for the more general $\kappa\gamma$-vacuum, whose Wightman function contains non-stationary, phase-dependent terms. It is an open question what physical system---perhaps a mirror with time-dependent reflectivity or a different boundary condition---could generate these richer correlation structures. Additionally, exploring detector trajectories that are not inertial could provide a direct probe of the non-stationary, local components of the Wightman functions analyzed in this paper. Finally, extending these ideas beyond $(1+1)$ dimensions remains a key challenge, where understanding the relationship between kinematic vacua and dynamical boundaries could yield new insights into the nature of horizons and quantum thermal effects in more realistic settings.

\section*{Acknowledgments}
I am grateful to Marlan Scully and Bill Unruh for discussions.  
This work was supported by the Robert A. Welch Foundation (Grant No. A-1261) and the National Science Foundation (Grant No. PHY-2013771).

\appendix

\section{\texorpdfstring{Mirror Worldline and Kinematics}{Mirror Worldline and Kinematics}}
\label{app:tx-mirror}

This appendix provides the explicit relationship between the abstract ray-tracing function $v=p(u)$ and the physical worldline $x=z(t)$ of the mirror in Minkowski coordinates. We also derive general expressions for the mirror's kinematics.

The inverse map between null and Minkowski coordinates is
\begin{align}
t=\frac{u+v}{2},\qquad x=\frac{v-u}{2}.
\end{align}
A mirror's worldline is the curve $\{(u,p(u))\}$, which in $(t,x)$ coordinates is parameterized by $u$:
\begin{align}
t(u)=\frac{u+p(u)}{2},\qquad x(u)=\frac{p(u)-u}{2}.
\label{eq:t-x-param}
\end{align}
Differentiating with respect to $u$ gives the mirror’s coordinate velocity and Lorentz factor directly in terms of $p(u)$:
\begin{align}
\frac{dx}{dt}=\frac{p'(u)-1}{p'(u)+1}\;=:\;v_{\rm mir}(u),\qquad
\gamma(u)=\frac{1}{\sqrt{1-v_{\rm mir}^2(u)}}=\frac{p'(u)+1}{2\sqrt{p'(u)}}.
\label{eq:v-gamma-general}
\end{align}
The proper acceleration is
\begin{align}
\alpha(u)=\gamma(u)^{3}\,\frac{d v_{\rm mir}}{dt} = \gamma(u)^{3}\,\frac{d v_{\rm mir}/du}{dt/du}
=\frac{p''(u)}{2\,[p'(u)]^{3/2}}.
\label{eq:proper-accel-general}
\end{align}
To describe the worldline as $x=z(t)$, one can eliminate $u$ via the implicit equation
\begin{align}
t+z(t)=p\big(t-z(t)\big).
\label{eq:zt-implicit}
\end{align}
For the Carlitz–Willey (CW) trajectory, we use the specific ray-tracing map
\begin{align}
p(u)=v_H-\frac{1}{\kappa}\,e^{-\kappa u}\qquad(\kappa>0),
\label{eq:CW-map}
\end{align}
where $v_H$ is the future event horizon. Solving \eqref{eq:zt-implicit} gives a closed form for the worldline $z(t)$ in terms of the Lambert $W$ function:
\begin{empheq}[box=\widefbox]{align}
z(t)=v_H-t-\frac{1}{\kappa}\,
W\!\Big(e^{-\kappa(2t-v_H)}\Big).
\label{eq:CW-z-of-t}
\end{empheq}
The kinematics for the CW mirror are particularly simple because $p'(u)=e^{-\kappa u}$. Inserting this into the general relations above yields:
\begin{empheq}[box=\widefbox]{align}
v_{\rm mir}(u)=\frac{e^{-\kappa u}-1}{e^{-\kappa u}+1}
=-\tanh\!\Big(\frac{\kappa u}{2}\Big),\qquad
\gamma(u)=\cosh\!\Big(\frac{\kappa u}{2}\Big).
\label{eq:CW-vel-gamma}
\end{empheq}
The proper acceleration is then
\begin{empheq}[box=\widefbox]{align}
\alpha(u)
=-\frac{\kappa}{2}\,e^{\kappa u/2},
\label{eq:CW-proper-accel}
\end{empheq}
whose magnitude grows exponentially as the mirror asymptotes to the null line $t+x=v_H$.

\section{\texorpdfstring{Mode Normalization and Carlitz--Willey Kernels}{Mode Normalization and Carlitz--Willey Kernels}}
\label{app:inner-product}

Throughout we work in $(1+1)$ Minkowski spacetime with null coordinates
\begin{align}
u = t-x,
\qquad
v = t+x.
\end{align}
The mirror follows a smooth, strictly increasing trajectory
\begin{align}
v = p(u),
\qquad p'(u)>0,
\end{align}
with inverse ray–tracing function
\begin{align}
u = f(v) = p^{-1}(v).
\end{align}

The normalized ``in'' and ``out'' modes supported on the right of the mirror are
\begin{align}
\phi^{\rm in}_{\omega'}(u,v)
&=\frac{1}{\sqrt{4\pi\omega'}}\Big(e^{-i\omega' v}-e^{-i\omega' p(u)}\Big),
\label{eq:app-modes-in}\\
\phi^{\rm out}_{\omega}(u,v)
&=\frac{1}{\sqrt{4\pi\omega}}\Big(e^{-i\omega f(v)}-e^{-i\omega u}\Big),
\label{eq:app-modes-out}
\end{align}
which satisfy the Dirichlet condition $\phi|_{v=p(u)}=0=\phi|_{u=f(v)}$ and are defined for strictly positive frequencies $\omega,\omega'>0$.

On a null Cauchy surface
\begin{align}
\mathscr C=\{u\in\mathbb R\}\cup\{v\in\mathbb R\},
\end{align}
the Klein–Gordon inner product is
\begin{align}
\langle f,g\rangle
= i\!\int_{-\infty}^{+\infty}\!du\;\big(f^*\partial_u g-(\partial_u f^*)g\big)
+ i\!\int_{-\infty}^{+\infty}\!dv\;\big(f^*\partial_v g-(\partial_v f^*)g\big).
\label{eq:app-KG}
\end{align}

\subsection{\texorpdfstring{Normalization on \(\mathscr I^-\)}{Normalization on I-}}

Because \eqref{eq:app-KG} is conserved, we may evaluate it on any convenient Cauchy surface. A particularly simple choice is past null infinity,
\[
\mathscr I^- = \mathscr I^-_{R} \cup \mathscr I^-_{L},
\]
with
\[
\mathscr I^-_{R}:\; u = u_0\to -\infty,\ v\in\mathbb R,\qquad
\mathscr I^-_{L}:\; v = v_0\to -\infty,\ u\in\mathbb R.
\]
On $\mathscr I^-_{R}$ the affine parameter is $v$ and the flux is generated by $\partial_v$.

The ``in'' modes \eqref{eq:app-modes-in} are chosen to describe incoming right–moving waves, so they have support only on $\mathscr I^-_{R}$ and vanish on $\mathscr I^-_{L}$. For products of in–modes the Klein–Gordon product therefore reduces to
\begin{align}
\langle f,g\rangle
= i\!\int_{\mathscr I^-_{R}}\!dv\,
\big(f^*\,\partial_v g-(\partial_v f^*)\,g\big).
\end{align}
Restricting \eqref{eq:app-modes-in} to $\mathscr I^-_{R}$ by fixing $u=u_0\to -\infty$ gives
\begin{align}
\phi^{\rm in}_{\omega}(u_0,v)
=\frac{1}{\sqrt{4\pi\omega}}\Big(e^{-i\omega v}-e^{-i\omega p(u_0)}\Big),
\end{align}
where $p(u_0)$ is a constant independent of $v$. For two modes with frequencies $\omega,\omega'>0$,
\begin{align}
\partial_v \phi^{\rm in}_{\omega}(u_0,v)
= -\frac{i\omega}{\sqrt{4\pi\omega}}\,e^{-i\omega v},\qquad
\partial_v \phi^{\rm in *}_{\omega'}(u_0,v)
= \frac{i\omega'}{\sqrt{4\pi\omega'}}\,e^{+i\omega' v}.
\end{align}
Substituting into \eqref{eq:app-KG},
\begin{align}
\langle \phi^{\rm in}_{\omega'},\phi^{\rm in}_{\omega}\rangle
&= i\!\int_{-\infty}^{+\infty}\!dv\,
\Big(\phi^{\rm in *}_{\omega'}\partial_v\phi^{\rm in}_{\omega}
-(\partial_v\phi^{\rm in *}_{\omega'})\phi^{\rm in}_{\omega}\Big)\nn\\
&=\frac{1}{4\pi\sqrt{\omega\omega'}}\int_{-\infty}^{+\infty}\!dv\,
\Big[(\omega+\omega')\,e^{i(\omega'-\omega)v}
-\omega\,e^{i\omega' v}e^{-i\omega p(u_0)}
-\omega'\,e^{-i\omega v}e^{+i\omega' p(u_0)}\Big].
\label{eq:app-inin-expanded}
\end{align}
The $v$–integrals are standard Fourier transforms. For $\omega,\omega'>0$ the terms proportional to the constant reflection phase yield distributions $\propto\delta(\omega)$ or $\delta\omega'$ and vanish, so only the overlap of the propagating pieces survives:
\begin{align}
\int_{-\infty}^{+\infty}\!dv\ e^{i(\omega'-\omega)v}
=2\pi\,\delta(\omega'-\omega).
\end{align}
Inserting into \eqref{eq:app-inin-expanded},
\begin{align}
\langle \phi^{\rm in}_{\omega'},\phi^{\rm in}_{\omega}\rangle
=\frac{1}{4\pi\sqrt{\omega\omega'}}\big[2\pi(\omega+\omega')\delta(\omega'-\omega)\big]=\delta(\omega-\omega'),
\label{eq:app-inin-orth}
\end{align}
where in the last line we use $\omega'=\omega$ on the support of the delta function. Thus the in–modes are orthonormal on $\mathscr I^-_{R}$ and are Klein–Gordon–equivalent there to ordinary Minkowski plane waves in $v$.

\subsection{\texorpdfstring{Normalization on \(\mathscr I^+\)}{Normalization on I+}}

An entirely analogous argument holds on future null infinity,
\[
\mathscr I^+ = \mathscr I^+_{R} \cup \mathscr I^+_{L},
\]
with
\[
\mathscr I^+_{R}:\; v = v_0\to +\infty,\ u\in\mathbb R,\qquad
\mathscr I^+_{L}:\; u = u_0\to +\infty,\ v\in\mathbb R.
\]
The outgoing right–moving out–modes \eqref{eq:app-modes-out} have support only on $\mathscr I^+_{R}$, where the affine parameter is $u$ and the flux is generated by $\partial_u$. On this surface $v$ is fixed, so any term of the form $e^{-i\omega f(v)}$ becomes a $u$–independent constant and contributes only $\propto\omega\,\delta(\omega)$ or $\propto\omega'\,\delta\omega'$ to the Klein–Gordon product; these vanish for $\omega,\omega'>0$. Repeating the calculation with $u$ in place of $v$ yields
\begin{align}
\langle \phi^{\rm out}_{\omega'},\phi^{\rm out}_{\omega}\rangle
=\delta(\omega-\omega'),\qquad
\langle \phi^{\rm out}_{\omega'},\phi^{\rm out\,*}_{\omega}\rangle=0.
\end{align}
Combined with \eqref{eq:app-inin-orth}, this establishes
\begin{align}
\langle \phi^{\rm in}_{\omega'},\phi^{\rm in}_{\omega}\rangle
=\langle \phi^{\rm out}_{\omega'},\phi^{\rm out}_{\omega}\rangle
=\delta(\omega-\omega'),\qquad
\langle \phi^{\rm in}_{\omega'},\phi^{\rm in\,*}_{\omega}\rangle
=\langle \phi^{\rm out}_{\omega'},\phi^{\rm out\,*}_{\omega}\rangle=0,
\end{align}
independently of the detailed form of $p(u)$, including the Carlitz--Willey trajectory.

\subsection{\texorpdfstring{Bogoliubov Kernels for a General Mirror}{Bogoliubov Kernels for a General Mirror}}
\label{app:bogol-kernel}

Expanding the field in the normalized in– and out–modes,
\begin{align}
\Phi(u,v)
&=\int_{0}^{\infty}\!d\omega'\Big(
a^{\rm in}_{\omega'}\phi^{\rm in}_{\omega'}(u,v)
+a^{\rm in\,\dagger}_{\omega'}\phi^{\rm in *}_{\omega'}(u,v)\Big)\nn\\
&=\int_{0}^{\infty}\!d\omega\Big(
a^{\rm out}_{\omega}\phi^{\rm out}_{\omega}(u,v)
+a^{\rm out\,\dagger}_{\omega}\phi^{\rm out *}_{\omega}(u,v)\Big),
\end{align}
and using the orthonormality above, the Bogoliubov transformation
\begin{align}
a^{\rm out}_{\omega}
=\int_{0}^{\infty}\!d\omega'\,\Big(
\alpha_{\omega\omega'}\,a^{\rm in}_{\omega'}
+\beta_{\omega\omega'}\,a^{\rm in\,\dagger}_{\omega'}\Big)
\end{align}
has kernels
\begin{align}
\alpha_{\omega\omega'}
&=\langle \phi^{\rm out}_{\omega},\phi^{\rm in}_{\omega'}\rangle_{\mathscr I^-_R}
= i\!\int_{-\infty}^{+\infty}\!dv\,
\Big(\phi^{\rm out\,*}_{\omega}\partial_v\phi^{\rm in}_{\omega'}
-(\partial_v\phi^{\rm out\,*}_{\omega})\phi^{\rm in}_{\omega'}\Big),\\
\beta_{\omega\omega'}
&=-\langle \phi^{\rm out}_{\omega},\phi^{\rm in\,*}_{\omega'}\rangle_{\mathscr I^-_R}.
\end{align}
On $\mathscr I^-_R$ we fix $u=u_0\to-\infty$ and let $v\in\mathbb R$. As before, the constant reflection pieces drop out and only the dynamical phases contribute:
\begin{align}
\partial_v\phi^{\rm in}_{\omega'} &\sim -i\omega' e^{-i\omega' v},\qquad
\partial_v\phi^{\rm out\,*}_{\omega} \sim i\omega f'(v)e^{i\omega f(v)}.
\end{align}
Substituting,
\begin{align}
\alpha_{\omega\omega'}
&=\frac{1}{4\pi\sqrt{\omega\omega'}}\!
\int_{-\infty}^{+\infty}\!dv\,
\big(\omega'+\omega f'(v)\big)e^{i\omega f(v)}e^{-i\omega' v}.
\label{eq:app-alpha-raw-again}
\end{align}
Using
\begin{align}
\partial_v\!\left(e^{i\omega f(v)}e^{-i\omega' v}\right)
=i\big(\omega f'(v)-\omega'\big)e^{i\omega f(v)}e^{-i\omega' v},
\end{align}
we can rewrite the integrand as
\begin{align}
\big(\omega'+\omega f'(v)\big)e^{i\omega f(v)}e^{-i\omega' v}
=2\omega' e^{i\omega f(v)}e^{-i\omega' v}
+\frac{1}{i}\,\partial_v\!\left(e^{i\omega f(v)}e^{-i\omega' v}\right).
\end{align}
For physically admissible trajectories (with an $i\epsilon$ prescription or wave–packet smearing) the total derivative gives a vanishing boundary term, and we obtain the standard integral form
\begin{align}
\boxed{
\alpha_{\omega\omega'}
=\frac{1}{2\pi}\sqrt{\frac{\omega'}{\omega}}
\int_{-\infty}^{+\infty}\!dv\,
e^{\,i\omega f(v)}\,e^{-i\omega' v}
}\,,
\label{eq:app-alpha-final}
\end{align}
while the $\beta$–kernel is
\begin{align}
\boxed{
\beta_{\omega\omega'}
=\frac{1}{2\pi}\sqrt{\frac{\omega'}{\omega}}
\int_{-\infty}^{+\infty}\!dv\,
e^{\,i\omega f(v)}\,e^{\,i\omega' v}
}\,.
\label{eq:app-beta-final}
\end{align}

\subsection{Carlitz--Willey Bogoliubov kernels}
\label{app:CW-kernels}

For the Carlitz--Willey trajectory we work, after an affine shift of $v$, with the ray-tracing map
\begin{align}
p(u)=-A\,e^{-\kappa u},
\qquad
A=\frac{1}{\kappa}>0,
\end{align}
so that the mirror asymptotes to the null line $v=0$ from below and $p(u)<0$ for all finite $u$.
The inverse ray-tracing function is
\begin{align}
v=p(u)=-A e^{-\kappa u}
\quad\Longrightarrow\quad
u=f(v)=-\frac{1}{\kappa}\ln\!\left(-\frac{v}{A}\right),
\label{eq:f-of-v-CW}
\end{align}
which is real only for $v<0$. Hence the integrals for the Bogoliubov kernels
 (\ref{eq:app-alpha-final}) and (\ref{eq:app-beta-final}) effectively reduce to $v\in(-\infty,0)$.

From \eqref{eq:f-of-v-CW} we obtain the phase factor
\begin{align}
e^{i\omega f(v)}
=e^{\,i\omega\left(-\frac{1}{\kappa}\ln\!\left(-\frac{v}{A}\right)\right)}
=\left(-\frac{v}{A}\right)^{-i\omega/\kappa}
=A^{\,i\omega/\kappa}(-v)^{-i\omega/\kappa}.
\label{eq:phase-factor-CW}
\end{align}
Thus
\begin{align}
\alpha_{\omega\omega'}
&=\frac{1}{2\pi}\sqrt{\frac{\omega'}{\omega}}\,A^{i\omega/\kappa}
\int_{-\infty}^{0}\!dv\,(-v)^{-i\omega/\kappa}\,e^{-i\omega' v},
\label{eq:alpha-pre-int}
\\
\beta_{\omega\omega'}
&=\frac{1}{2\pi}\sqrt{\frac{\omega'}{\omega}}\,A^{i\omega/\kappa}
\int_{-\infty}^{0}\!dv\,(-v)^{-i\omega/\kappa}\,e^{+i\omega' v}.
\label{eq:beta-pre-int}
\end{align}


For the ``$\alpha$'' kernel we set $x=-v>0$, so that $v=-x$, $dv=-dx$ and $-v=x$. Then
\begin{align}
\int_{-\infty}^{0}\!dv\,(-v)^{-i\omega/\kappa}e^{-i\omega'v}
&=\int_{+\infty}^{0}\!(-dx)\,x^{-i\omega/\kappa}e^{i\omega' x}
=\int_{0}^{\infty}\!dx\,x^{-i\omega/\kappa}e^{i\omega' x}.
\end{align}
Write $x^{-i\omega/\kappa}=x^{s-1}$ with
\begin{align}
s=1-\frac{i\omega}{\kappa},
\end{align}
and use the standard Gamma–function representation
\begin{align}
\int_{0}^{\infty}\!dx\,x^{s-1}e^{-b x}
=b^{-s}\Gamma(s),
\qquad \Re(s)>0,\ \Re(b)>0.
\end{align}
To represent $e^{i\omega' x}$ we choose
\begin{align}
b=-i\omega'+\epsilon,
\qquad \epsilon>0,
\end{align}
so that $e^{-b x}=e^{i\omega' x-\epsilon x}$ and the integral is defined by an $i\epsilon$ prescription. Hence
\begin{align}
\int_{0}^{\infty}\!dx\,x^{-i\omega/\kappa}e^{i\omega' x}
=b^{-s}\Gamma(s),
\qquad
s=1-\frac{i\omega}{\kappa},\quad b=-i\omega'+\epsilon.
\end{align}
On the principal branch we have
\begin{align}
\ln(-i\omega'+\epsilon)
=\ln\omega'-i\frac{\pi}{2},
\end{align}
so
\begin{align}
b^{-s}
&=\exp[-s\ln b]
=\exp\!\left[-\Big(1-\frac{i\omega}{\kappa}\Big)\Big(\ln\omega'-i\frac{\pi}{2}\Big)\right]\nn\\
&=\omega'^{-\left(1-\frac{i\omega}{\kappa}\right)}\,
\exp\!\left(i\frac{\pi}{2}\right)\,
\exp\!\left(\frac{\pi\omega}{2\kappa}\right)\nn\\
&=\omega'^{-\left(1-\frac{i\omega}{\kappa}\right)}\,
i\,e^{\frac{\pi\omega}{2\kappa}}.
\end{align}
Therefore
\begin{align}
\int_{0}^{\infty}\!dx\,x^{-i\omega/\kappa}e^{i\omega' x}
&=i\,\omega'^{-\left(1-\frac{i\omega}{\kappa}\right)}
e^{\frac{\pi\omega}{2\kappa}}
\Gamma\!\left(1-\frac{i\omega}{\kappa}\right),
\end{align}
and \eqref{eq:alpha-pre-int} becomes
\begin{align}
\alpha_{\omega\omega'}
&=\frac{1}{2\pi}\sqrt{\frac{\omega'}{\omega}}\,A^{i\omega/\kappa}\,
i\,\omega'^{-\left(1-\frac{i\omega}{\kappa}\right)}
e^{\frac{\pi\omega}{2\kappa}}
\Gamma\!\left(1-\frac{i\omega}{\kappa}\right)\nn\\
&=\frac{i}{2\pi}\frac{1}{\sqrt{\omega}}\,
A^{i\omega/\kappa}\,
e^{\frac{\pi\omega}{2\kappa}}\,
\omega'^{-1/2+i\omega/\kappa}\,
\Gamma\!\left(1-\frac{i\omega}{\kappa}\right)\nn\\
&=\frac{i}{2\pi}\frac{1}{\sqrt{\omega\omega'}}\,
(\omega' A)^{i\omega/\kappa}\,
e^{\frac{\pi\omega}{2\kappa}}\,
\Gamma\!\left(1-\frac{i\omega}{\kappa}\right).
\label{eq:alpha-omega-omega-prime}
\end{align}


For the ``$\beta$'' kernel the same change of variables gives
\begin{align}
\int_{-\infty}^{0}\!dv\,(-v)^{-i\omega/\kappa}e^{+i\omega'v}
&=\int_{0}^{\infty}\!dx\,x^{-i\omega/\kappa}e^{-i\omega' x}.
\end{align}
Again write $x^{-i\omega/\kappa}=x^{s-1}$ with $s=1-i\omega/\kappa$, and now choose
\begin{align}
b=i\omega'+\epsilon,
\end{align}
so that $e^{-b x}=e^{-i\omega' x-\epsilon x}$ and
\begin{align}
\int_{0}^{\infty}\!dx\,x^{-i\omega/\kappa}e^{-i\omega' x}
=b^{-s}\Gamma(s),
\qquad s=1-\frac{i\omega}{\kappa},\quad b=i\omega'+\epsilon.
\end{align}
On the principal branch
\begin{align}
\ln(i\omega'+\epsilon)
=\ln\omega'+i\frac{\pi}{2},
\end{align}
so
\begin{align}
b^{-s}
&=\exp[-s\ln b]
=\exp\!\left[-\Big(1-\frac{i\omega}{\kappa}\Big)\Big(\ln\omega'+i\frac{\pi}{2}\Big)\right]\nn\\
&=\omega'^{-\left(1-\frac{i\omega}{\kappa}\right)}\,
\exp\!\left(-i\frac{\pi}{2}\right)\,
\exp\!\left(-\frac{\pi\omega}{2\kappa}\right)\nn\\
&=\omega'^{-\left(1-\frac{i\omega}{\kappa}\right)}\,
(-i)\,e^{-\frac{\pi\omega}{2\kappa}}.
\end{align}
Hence
\begin{align}
\int_{0}^{\infty}\!dx\,x^{-i\omega/\kappa}e^{-i\omega' x}
&=(-i)\,\omega'^{-\left(1-\frac{i\omega}{\kappa}\right)}
e^{-\frac{\pi\omega}{2\kappa}}
\Gamma\!\left(1-\frac{i\omega}{\kappa}\right),
\end{align}
and \eqref{eq:beta-pre-int} gives
\begin{align}
\beta_{\omega\omega'}
&=\frac{1}{2\pi}\sqrt{\frac{\omega'}{\omega}}\,A^{i\omega/\kappa}\,
(-i)\,\omega'^{-\left(1-\frac{i\omega}{\kappa}\right)}
e^{-\frac{\pi\omega}{2\kappa}}
\Gamma\!\left(1-\frac{i\omega}{\kappa}\right)\nn\\
&=\frac{-i}{2\pi}\frac{1}{\sqrt{\omega}}\,
A^{i\omega/\kappa}\,
e^{-\frac{\pi\omega}{2\kappa}}\,
\omega'^{-1/2+i\omega/\kappa}\,
\Gamma\!\left(1-\frac{i\omega}{\kappa}\right)\nn\\
&=\frac{-i}{2\pi}\frac{1}{\sqrt{\omega\omega'}}\,
(\omega' A)^{i\omega/\kappa}\,
e^{-\frac{\pi\omega}{2\kappa}}\,
\Gamma\!\left(1-\frac{i\omega}{\kappa}\right).
\label{eq:beta-omega-omega-prime}
\end{align}
Taking absolute values, one finds the thermal ratio
\begin{align}
\frac{|\beta_{\omega\omega'}|^{2}}{|\alpha_{\omega\omega'}|^{2}}
=e^{-2\pi\omega/\kappa},
\end{align}
corresponding to a Planck spectrum at temperature $T_{\rm CW}=\kappa/(2\pi)$.

\section{\texorpdfstring{Mellin Basis and Single-Mode Bogoliubov Map}{Mellin Basis and Single-Mode Bogoliubov Map}}
\label{app:Mellin-Bogol}

In this appendix we show how a Mellin transform in the \emph{in}–frequency
diagonalizes the Carlitz--Willey Bogoliubov kernels
\eqref{eq:alpha-omega-omega-prime}–\eqref{eq:beta-omega-omega-prime},
so that each out–frequency $\omega$ couples only to the Mellin labels
$(\omega,-\omega)$ of the in–sector. In this sense the Carlitz--Willey
mirror acts as a ``single–mode'' SU(1,1) transformation, with no sum over
different frequencies.

\subsection{\texorpdfstring{Mellin Transform of In–Operators}{Mellin Transform of In–Operators}}

We introduce the Mellin–transformed in–operators
\begin{align}
b^{\rm in}_{\Omega}
=\int_{0}^{\infty}\frac{d\omega'}{\sqrt{2\pi\kappa}}\,
\omega'^{-\frac{1}{2}-i\Omega/\kappa}\,a^{\rm in}_{\omega'},
\qquad \Omega\in\mathbb{R},
\label{eq:Mellin-in-def}
\end{align}
with inverse transform
\begin{align}
a^{\rm in}_{\omega'}
=\int_{-\infty}^{+\infty}\frac{d\Omega}{\sqrt{2\pi\kappa}}\,
\omega'^{-\frac{1}{2}+i\Omega/\kappa}\,b^{\rm in}_{\Omega}.
\label{eq:Mellin-in-inverse}
\end{align}
The inversion formula follows from the standard Mellin orthogonality
\begin{align}
\int_{0}^{\infty}\frac{d\omega'}{2\pi\kappa}\,
\omega'^{-1+i(\Omega-\Omega')/\kappa}
=\delta(\Omega-\Omega'),
\label{eq:Mellin-orthogonality}
\end{align}
which is obtained by the logarithmic change of variables
$\omega'=e^{x}$, $d\omega'=e^{x}dx$, giving
\begin{align}
\int_{0}^{\infty}\frac{d\omega'}{2\pi\kappa}\,
\omega'^{-1+i(\Omega-\Omega')/\kappa}
&=\frac{1}{2\pi\kappa}\int_{-\infty}^{+\infty}dx\,
e^{i(\Omega-\Omega')x/\kappa}
=\delta(\Omega-\Omega').
\end{align}

Using $[a^{\rm in}_{\omega},a^{\rm in\,\dagger}_{\omega'}]=\delta(\omega-\omega')$,
one finds
\begin{align}
[b^{\rm in}_{\Omega},b^{\rm in\,\dagger}_{\Omega'}]
=\delta(\Omega-\Omega'),
\end{align}
so the $b^{\rm in}_{\Omega}$ form a valid bosonic basis. It is convenient
for the algebra to keep $\Omega\in\mathbb{R}$; one may later reorganize
the $\pm\Omega$ sectors if desired.

\subsection{\texorpdfstring{Diagonalization of the CW Kernels}{Diagonalization of the CW Kernels}}

The Carlitz--Willey Bogoliubov kernels
\eqref{eq:alpha-omega-omega-prime}–\eqref{eq:beta-omega-omega-prime}
have the factorized form
\begin{align}
\alpha_{\omega\omega'}
&=F(\omega)\,\omega'^{-1/2+i\omega/\kappa},\qquad
\beta_{\omega\omega'}
=G(\omega)\,\omega'^{-1/2+i\omega/\kappa},
\label{eq:alpha-beta-factorized}
\end{align}
with
\begin{align}
F(\omega)
&=\frac{i}{2\pi}\sqrt{\frac{1}{\omega}}\,
A^{i\omega/\kappa}\,e^{+\pi\omega/(2\kappa)}\,
\Gamma\!\left(1-\frac{i\omega}{\kappa}\right),\\[1mm]
G(\omega)
&=\frac{-i}{2\pi}\sqrt{\frac{1}{\omega}}\,
A^{i\omega/\kappa}\,e^{-\pi\omega/(2\kappa)}\,
\Gamma\!\left(1-\frac{i\omega}{\kappa}\right),
\label{eq:F-G-def}
\end{align}
where the overall phases $\pm i$ are kept explicitly, as in
\eqref{eq:alpha-omega-omega-prime}–\eqref{eq:beta-omega-omega-prime}.
All the $\omega'$–dependence resides in the Mellin factor
$\omega'^{-1/2+i\omega/\kappa}$, so the Mellin transform in $\omega'$
diagonalizes the Bogoliubov map.

Start from the usual Bogoliubov relation
\begin{align}
a^{\rm out}_{\omega}
=\int_{0}^{\infty}\!d\omega'\,
\Big(\alpha_{\omega\omega'}\,a^{\rm in}_{\omega'}
+\beta_{\omega\omega'}\,a^{\rm in\,\dagger}_{\omega'}\Big).
\label{eq:usual-Bogol}
\end{align}
Insert the inverse Mellin transform \eqref{eq:Mellin-in-inverse} and its
Hermitian conjugate:
\begin{align}
a^{\rm in}_{\omega'}
&=\int_{-\infty}^{+\infty}\frac{d\Omega}{\sqrt{2\pi\kappa}}\,
\omega'^{-1/2+i\Omega/\kappa}\,b^{\rm in}_{\Omega},\\
a^{\rm in\,\dagger}_{\omega'}
&=\int_{-\infty}^{+\infty}\frac{d\Omega}{\sqrt{2\pi\kappa}}\,
\omega'^{-1/2-i\Omega/\kappa}\,b^{\rm in\,\dagger}_{\Omega}.
\end{align}
Using \eqref{eq:alpha-beta-factorized}, we obtain
\begin{align}
a^{\rm out}_{\omega}
&=\int_{0}^{\infty}\!d\omega'\,F(\omega)\,\omega'^{-1/2+i\omega/\kappa}
\int_{-\infty}^{+\infty}\frac{d\Omega}{\sqrt{2\pi\kappa}}\,
\omega'^{-1/2+i\Omega/\kappa}\,b^{\rm in}_{\Omega}\nn\\
&\quad+\int_{0}^{\infty}\!d\omega'\,G(\omega)\,\omega'^{-1/2+i\omega/\kappa}
\int_{-\infty}^{+\infty}\frac{d\Omega}{\sqrt{2\pi\kappa}}\,
\omega'^{-1/2-i\Omega/\kappa}\,b^{\rm in\,\dagger}_{\Omega}.
\end{align}
Interchanging the $\omega'$ and $\Omega$ integrals and applying
\eqref{eq:Mellin-orthogonality},
\begin{align}
\int_{0}^{\infty}\frac{d\omega'}{2\pi\kappa}\,
\omega'^{-1+i(\omega-\Omega)/\kappa}
&=\delta(\omega-\Omega),\\
\int_{0}^{\infty}\frac{d\omega'}{2\pi\kappa}\,
\omega'^{-1+i(\omega+\Omega)/\kappa}
&=\delta(\omega+\Omega),
\end{align}
we find
\begin{align}
a^{\rm out}_{\omega}
&=F(\omega)\int_{-\infty}^{+\infty}\!d\Omega\,
\delta(\omega-\Omega)\,b^{\rm in}_{\Omega}
+G(\omega)\int_{-\infty}^{+\infty}\!d\Omega\,
\delta(\omega+\Omega)\,b^{\rm in\,\dagger}_{\Omega}\nn\\
&=F(\omega)\,b^{\rm in}_{\omega}
+G(\omega)\,b^{\rm in\,\dagger}_{-\omega}.
\label{eq:aout-in-basis}
\end{align}

Equation \eqref{eq:aout-in-basis} is the central result: the original
double integral over $\omega'$ in \eqref{eq:usual-Bogol} has collapsed to
an SU(1,1)–type transformation acting only on the pair of Mellin labels
$(\omega,-\omega)$, with no integral over any other frequency. The Mellin
spectrum $\Omega\in\mathbb{R}$ naturally appears in $\pm$ pairs, and each
out–frequency $\omega$ couples only to the corresponding pair
$(\omega,-\omega)$ in the in–sector.

\subsection{\texorpdfstring{Single-Mode SU(1,1) Structure and Thermal Ratio}{Single-Mode SU(1,1) Structure and Thermal Ratio}}

For each real label $\Omega$ the pair
\begin{align}
\big(b^{\rm in}_{\Omega},\,b^{\rm in\,\dagger}_{-\Omega}\big)
\end{align}
spans a single canonical bosonic degree of freedom. Equation
\eqref{eq:aout-in-basis} shows that, for $\omega=\Omega$, the out–operator
acts within this two–dimensional subspace as
\begin{align}
a^{\rm out}_{\Omega}
=F(\Omega)\,b^{\rm in}_{\Omega}
+G(\Omega)\,b^{\rm in\,\dagger}_{-\Omega},
\qquad \Omega\in\mathbb{R},
\label{eq:single-mode-SU11}
\end{align}
with $F$ and $G$ given explicitly by \eqref{eq:F-G-def}. There is no sum
over different Mellin labels: each $(\Omega,-\Omega)$ block undergoes an
independent SU(1,1) Bogoliubov transformation.

The ratio of coefficients follows immediately from \eqref{eq:F-G-def}:
\begin{align}
\frac{G(\Omega)}{F(\Omega)}
&=\frac{-i\,A^{i\Omega/\kappa}e^{-\pi\Omega/(2\kappa)}
\Gamma\!\left(1-\frac{i\Omega}{\kappa}\right)}
{i\,A^{i\Omega/\kappa}e^{+\pi\Omega/(2\kappa)}
\Gamma\!\left(1-\frac{i\Omega}{\kappa}\right)}
=e^{-\pi\Omega/\kappa},
\end{align}
so that
\begin{align}
\frac{|G(\Omega)|^{2}}{|F(\Omega)|^{2}}
=e^{-2\pi\Omega/\kappa}.
\end{align}
Up to an overall, frequency–dependent normalization of the out–basis
(which can be absorbed into a redefinition of $a^{\rm out}_{\Omega}$ so
that $|F(\Omega)|^{2}-|G(\Omega)|^{2}=1$), this is precisely the SU(1,1)
relation for a single squeezed mode. The mean occupation number in the
in–vacuum then takes the Planck form
\begin{align}
\langle 0_{\rm in}|\,
a^{\rm out\,\dagger}_{\Omega}a^{\rm out}_{\Omega}\,
|0_{\rm in}\rangle
=|G(\Omega)|^{2}
=\frac{1}{e^{2\pi\Omega/\kappa}-1},
\end{align}
demonstrating explicitly that, in the Mellin basis, the Carlitz--Willey
mirror acts as a collection of independent single–mode SU(1,1)
transformations, one for each pair $(\Omega,-\Omega)$, with thermal ratio
$|G(\Omega)|^{2}/|F(\Omega)|^{2}=e^{-2\pi\Omega/\kappa}$ and no sum over
different frequencies.

\section{\texorpdfstring{Proof of the Commutation Relation for Mellin-Transformed Operators}{Proof of Commutation Relation for Mellin Operators}}
\label{app:Mellin}

We wish to prove that if the standard ``in''-mode operators obey the canonical commutation relation
\begin{align}
[a^{\rm in}_{\omega}, a^{\rm in \dagger}_{\omega'}] = \delta(\omega-\omega'),
\end{align}
then the Mellin-transformed operators, defined as
\begin{align}
b_{\Omega}
=\int_{0}^{\infty}\frac{d\omega'}{\sqrt{2\pi\kappa}}\,
\omega'^{-\frac12 - i\Omega/\kappa}\,a^{\rm in}_{\omega'},
\end{align}
also obey the canonical commutation relation $[b_{\Omega}, b_{\Omega'}^{\dagger}]=\delta(\Omega-\Omega')$.

\begin{proof}
We first write out the commutator using the definitions of $b_{\Omega}$ and its Hermitian conjugate $b_{\Omega'}^{\dagger}$:
\begin{align}
[b_{\Omega}, b_{\Omega'}^{\dagger}]
= \left[ \int_{0}^{\infty}\frac{d\omega}{\sqrt{2\pi\kappa}}\, \omega^{-\frac12 - i\Omega/\kappa}\,a^{\rm in}_{\omega} \ , \
\int_{0}^{\infty}\frac{d\omega'}{\sqrt{2\pi\kappa}}\, \omega'^{-\frac12 + i\Omega'/\kappa}\,a^{\rm in \dagger}_{\omega'} \right].
\end{align}
Using the linearity of the commutator and substituting the canonical relation for the $a^{\rm in}_{\omega}$,
\begin{align}
[b_{\Omega}, b_{\Omega'}^{\dagger}]
&= \frac{1}{2\pi\kappa} \int_{0}^{\infty} d\omega \int_{0}^{\infty} d\omega' \,
\omega^{-\frac12 - i\Omega/\kappa} \omega'^{-\frac12 + i\Omega'/\kappa}
[a^{\rm in}_{\omega}, a^{\rm in \dagger}_{\omega'}] \nonumber \\
&= \frac{1}{2\pi\kappa} \int_{0}^{\infty} d\omega \int_{0}^{\infty} d\omega' \,
\omega^{-\frac12 - i\Omega/\kappa} \omega'^{-\frac12 + i\Omega'/\kappa}
\delta(\omega-\omega').
\end{align}
The Dirac delta collapses the $\omega'$-integral, giving
\begin{align}
[b_{\Omega}, b_{\Omega'}^{\dagger}]
= \frac{1}{2\pi\kappa} \int_{0}^{\infty} d\omega \,
\omega^{-1 - i(\Omega-\Omega')/\kappa}.
\end{align}
This integral is evaluated most conveniently by a logarithmic change of variables. Let $\omega = e^x$, so $d\omega = e^x dx$ and the limits $(\omega=0,\infty)$ map to $(x=-\infty,\infty)$. The integrand becomes
\begin{align}
\omega^{-1 - i(\Omega-\Omega')/\kappa}
= e^{-x} e^{-i(\Omega-\Omega')x/\kappa},
\end{align}
so
\begin{align}
\int_{0}^{\infty} d\omega \, \omega^{-1 - i(\Omega-\Omega')/\kappa}
&= \int_{-\infty}^{\infty} e^x dx \, e^{-x} e^{-i(\Omega-\Omega')x/\kappa}
\nonumber\\
&= \int_{-\infty}^{\infty} e^{-i\frac{(\Omega-\Omega')}{\kappa}x} \, dx.
\end{align}
This is the standard Fourier representation of the Dirac delta function,
\begin{align}
\int_{-\infty}^{\infty} e^{-ikx}\,dx = 2\pi\delta(k),
\end{align}
which here implies
\begin{align}
\int_{-\infty}^{\infty} e^{-i\frac{(\Omega-\Omega')}{\kappa}x} \, dx
= 2\pi\delta\!\left(\frac{\Omega-\Omega'}{\kappa}\right).
\end{align}
Substituting back into the commutator and using the scaling property $\delta(a x) = |a|^{-1}\delta(x)$, we obtain (for $\kappa>0$)
\begin{align}
[b_{\Omega}, b_{\Omega'}^{\dagger}]
&= \frac{1}{2\pi\kappa}\left[2\pi\delta\!\left(\frac{\Omega-\Omega'}{\kappa}\right)\right]
= \frac{1}{\kappa}\,\kappa\,\delta(\Omega-\Omega')
= \delta(\Omega-\Omega').
\end{align}
This confirms that the Mellin-transformed operators satisfy the canonical bosonic commutation relations.
\end{proof}

\section{\texorpdfstring{Wightman Function Derivations}{Wightman Function Derivations}}
\label{app:Wightman_derivations}
This appendix provides the detailed derivations for the Wightman functions used in the main text.
\subsection{\texorpdfstring{The $\kappa$-plane-wave vacuum}{The kappa-plane-wave vacuum}}
\label{app:Wightman_kappa}
The Wightman function for the $\kappa$-plane-wave vacuum is
\begin{align}
W^{\mathrm{RTW}}_{\kappa}(u,u')
=\langle 0_{\kappa}\,|\,\Phi_{\mathrm{RTW}}(u)\,\Phi_{\mathrm{RTW}}(u')\,|\,0_{\kappa}\rangle.
\end{align}
Using the field expansion and $[{\cal A}_{\Lambda},{\cal A}_{\Lambda'}^{\dagger}]=\delta(\Lambda-\Lambda')$, this reduces to a single integral:
\begin{align}
W^{\mathrm{RTW}}_{\kappa}(u,u')
&= \int_{0}^{\infty}\!d\Lambda\,\Phi^{(\kappa)}_{\Lambda}(u)\,\Phi^{(\kappa)}_{\Lambda}(u')^{*}\nn\\
&= \int_{0}^{\infty}\!\frac{d\Lambda}{{\cal N}_{\Lambda,\kappa}}
\Big(e^{\frac{\pi\Lambda}{2\kappa}}e^{-i\Lambda u}+e^{-\frac{\pi\Lambda}{2\kappa}}e^{+i\Lambda u}\Big)
\Big(e^{\frac{\pi\Lambda}{2\kappa}}e^{+i\Lambda u'}+e^{-\frac{\pi\Lambda}{2\kappa}}e^{-i\Lambda u'}\Big),
\end{align}
with ${\cal N}_{\Lambda,\kappa}=8\pi\Lambda\,\sinh(\pi\Lambda/\kappa)$.
Expanding the product gives
\begin{align}
W^{\mathrm{RTW}}_{\kappa}(u,u')
=\int_{0}^{\infty}\!\frac{d\Lambda}{{\cal N}_{\Lambda,\kappa}}
\Big[e^{\frac{\pi\Lambda}{\kappa}}e^{-i\Lambda(u-u')}
+e^{-\frac{\pi\Lambda}{\kappa}}e^{+i\Lambda(u-u')}
+e^{-i\Lambda(u+u')}
+e^{+i\Lambda(u+u')}\Big].
\end{align}
Using
\begin{align}
e^{A-iB}+e^{-A+iB}=2\cosh(A-iB),\qquad
e^{-iC}+e^{iC}=2\cos C,
\end{align}
we can write this as
\begin{align}
W^{\mathrm{RTW}}_{\kappa}(u,u')
&=\int_{0}^{\infty}\!\frac{d\Lambda}{4\pi\Lambda\,\sinh\!\left(\frac{\pi\Lambda}{\kappa}\right)}
\,\cosh\!\Big(\frac{\pi\Lambda}{\kappa}-i\Lambda\Delta u\Big)\nn\\
&\quad+\int_{0}^{\infty}\!\frac{d\Lambda}{4\pi\Lambda\,\sinh\!\left(\frac{\pi\Lambda}{\kappa}\right)}\,
\cos\!\big(\Lambda(u+u')\big),
\end{align}
where $\Delta u:=u-u'$.

The second integral can be evaluated by differentiating with respect to $(u+u')$ and using the standard identity
\begin{align}
\int_{0}^{\infty}\!d\Lambda\,\frac{\sin(a\Lambda)}{\sinh(b\Lambda)}
= \frac{\pi}{2b}\tanh\!\left(\frac{\pi a}{2b}\right),
\qquad a>0,\;b>0,
\end{align}
see, e.g.,~\cite{Gradshteyn_Ryzhik2014}.
Integrating back and fixing an irrelevant additive constant yields
\begin{align}
\int_{0}^{\infty}\!\frac{d\Lambda}{\Lambda\,\sinh\!\left(\frac{\pi\Lambda}{\kappa}\right)}
\cos\!\big(\Lambda(u+u')\big)
= -\ln\!\cosh\!\Big(\frac{\kappa}{2}(u+u')\Big)+\text{const}.
\end{align}
A similar evaluation of the first integral gives the usual thermal expression in terms of $\sinh\!\big(\frac{\kappa}{2}(\Delta u-i\epsilon)\big)$~\cite{Birrell_Davies1982}.
Up to an overall additive constant (absorbed into the renormalization scale $\mu$), we obtain the stationary and non-stationary pieces quoted in the main text:
\begin{align}
W_{\mathrm{stationary}}^{\mathrm{RTW}}(\Delta u)
&=-\frac{1}{4\pi}\ln\!\Big[\mu^{2}\,\frac{2}{\kappa}\,
\sinh\!\Big(\frac{\kappa}{2}(\Delta u-i\epsilon)\Big)\Big],\\
W_{\mathrm{non\mbox{-}stationary}}^{\mathrm{RTW}}(u{+}u')
&=-\frac{1}{4\pi}\ln\!\cosh\!\Big(\frac{\kappa}{2}(u{+}u')\Big).
\end{align}
Their sum gives the full right-moving Wightman function quoted in the main text.
The $\Delta u$-dependence is KMS at temperature $T=\kappa/(2\pi)$; the $u{+}u'$ term encodes the local squeezing profile of the RTW state.
\subsection{\texorpdfstring{The Carlitz--Willey mirror}{The Carlitz--Willey mirror}}
Expanding the field in the ``in'' basis, $a_{\omega'}\ket{0_{\rm in}}=0$, the bulk Wightman function is the mode sum
\begin{align}
W^{>}(u,v;u',v')
&:=\bra{0_{\rm in}}\Phi(u,v)\,\Phi(u',v')\ket{0_{\rm in}}\nn\\
&=\int_{0}^{\infty}\!d\omega'\;\phi^{\rm in}_{\omega'}(u,v)\,\phi^{\rm in\,*}_{\omega'}(u',v')\nn\\
&=\frac{1}{4\pi}\int_{0}^{\infty}\!\frac{d\omega'}{\omega'}\Big[
e^{-i\omega'(v-v')}
- e^{-i\omega'(v-p(u'))}
- e^{-i\omega'(p(u)-v')}
+ e^{-i\omega'(p(u)-p(u'))}\Big],
\label{eq:app-Wmodesum}
\end{align}
where $\phi^{\rm in}_{\omega'}(u,v)=\frac{1}{\sqrt{4\pi\omega'}}\big(e^{-i\omega' v}-e^{-i\omega' p(u)}\big)$ on the right of the mirror.
Each term has the elementary form
\begin{align}
\int_{0}^{\infty}\!\frac{d\omega}{\omega}\,e^{-i\omega x}
=-\ln\!\big[\mu^{2}(x-i\epsilon)\big],
\end{align}
with the causal $i\epsilon$ understood; different choices of overall constant are absorbed into the same renormalization scale $\mu$.
Evaluating \eqref{eq:app-Wmodesum} then gives the standard image form for the bulk correlator:
\begin{align}
W^{>}(u,v;u',v')
=-\frac{1}{4\pi}\ln\!\left[
\frac{(v-v'-i\epsilon)\,\big(p(u)-p(u')-i\epsilon\big)}
{\big(v-p(u')-i\epsilon\big)\,\big(p(u)-v'-i\epsilon\big)}
\right],
\label{eq:app-Wimage}
\end{align}
up to an overall additive constant that is again absorbed into $\mu$.

For the chiral correlator of the reflected field on future null infinity, it is more convenient to proceed directly.
The general solution of the wave equation with a Dirichlet mirror at $v=p(u)$ can be written as
\begin{align}
\Phi(u,v) = \phi_{\rm in}(v) - \phi_{\rm in}\big(p(u)\big),
\end{align}
where $\phi_{\rm in}(v)$ is the incoming right-moving Minkowski field.
The outgoing chiral field on $\mathscr I^{+}$ (the $u$-line reached by rays that reflect off the mirror) is therefore
\begin{align}
\Phi_{\rm out}(u) := -\phi_{\rm in}\big(p(u)\big).
\end{align}
Its Wightman function in the ``in'' vacuum is
\begin{align}
W^{>}_{\rm out}(u,u')
&:=\bra{0_{\rm in}}\Phi_{\rm out}(u)\,\Phi_{\rm out}(u')\ket{0_{\rm in}}\nn\\
&=\bra{0_{\rm in}}\phi_{\rm in}\big(p(u)\big)\,\phi_{\rm in}\big(p(u')\big)\ket{0_{\rm in}}.
\end{align}
Using the standard chiral Minkowski correlator
\begin{align}
\bra{0_{\rm in}}\phi_{\rm in}(x)\,\phi_{\rm in}(x')\ket{0_{\rm in}}
=-\frac{1}{4\pi}\ln\!\big[\mu^{2}(x-x'-i\epsilon)\big],
\end{align}
we immediately obtain
\begin{align}
W^{>}_{\rm out}(u,u')
=-\frac{1}{4\pi}\,\ln\!\Big[\mu^{2}\,\big(p(u)-p(u')-i\epsilon\big)\Big].
\label{eq:W_chiral_Iplus}
\end{align}
For a static mirror (or no mirror) where $p(u)=u$, \eqref{eq:W_chiral_Iplus} reduces to the Minkowski right-moving correlator
\begin{align}
W^{>}_{\rm out}(u,u')=-\frac{1}{4\pi}\ln\!\big[\mu^{2}(u-u'-i\epsilon)\big],
\end{align}
as required.

\bibliographystyle{JHEP}
\bibliography{UnruhRef}

@article{Rindler66,
  author = "Rindler, W.",
  title = {{Kruskal Space and the Uniformly Accelerated Frame}},
  journal = "Am. J. Phys.",
  volume = "34",
  pages = "1174",
  year = "1966",
  doi = "10.1119/1.1972547"
}

@article{Unruh1976,
  title = {{Notes on black-hole evaporation}},
  author = {Unruh, W. G.},
  journal = {Phys. Rev. D},
  volume = {14},
  issue = {4},
  pages = {870--892},
  numpages = {0},
  year = {1976},
  month = {Aug},
  publisher = {American Physical Society},
  doi = {10.1103/PhysRevD.14.870},
  url = {https://link.aps.org/doi/10.1103/PhysRevD.14.870}
}

@article{Fulling1973,
  title = {{Nonuniqueness of Canonical Field Quantization in Riemannian Space-Time}},
  author = {Fulling, Stephen A.},
  journal = {Phys. Rev. D},
  volume = {7},
  issue = {10},
  pages = {2850--2862},
  numpages = {0},
  year = {1973},
  month = {May},
  publisher = {American Physical Society},
  doi = {10.1103/PhysRevD.7.2850},
  url = {https://link.aps.org/doi/10.1103/PhysRevD.7.2850}
}

@article{Hawking1975,
    author = "Hawking, S. W.",
    editor = "Gibbons, G. W. and Hawking, S. W.",
    title = "{Particle Creation by Black Holes}",
    doi = "10.1007/BF02345020",
    journal = "Commun. Math. Phys.",
    volume = "43",
    pages = "199--220",
    year = "1975",
    note = "[Erratum: Commun.Math.Phys. 46, 206 (1976)]"
}

@article{Davies1975,
doi = {10.1088/0305-4470/8/4/022},
url = {https://dx.doi.org/10.1088/0305-4470/8/4/022},
year = {1975},
month = {apr},
publisher = {},
volume = {8},
number = {4},
pages = {609},
author = {P C W Davies},
title = {{Scalar production in Schwarzschild and Rindler metrics}},
journal = {Journal of Physics A: Mathematical and General},
}

@book{Einstein100,
  author    = {DeWitt, Bryce S. },
  title     = {{General Relativity}: {An Einstein Centenary Survey}},
  isbn      = {978-0-521-29928-2},
  publisher = {Univ. Pr.},
  address   = {Cambridge, UK},
  year      = {1979}
}

@book{Birrell_Davies1982,
    author = "Birrell, N. D. and Davies, P. C. W.",
    title = "{Quantum Fields in Curved Space}",
    doi = "10.1017/CBO9780511622632",
    isbn = "978-0-511-62263-2, 978-0-521-27858-4",
    publisher = "Cambridge University Press",
    address = "Cambridge, UK",
    series = "Cambridge Monographs on Mathematical Physics",
    year = "1982"
}

@article{Colosi2009Rovelli,
doi = {10.1088/0264-9381/26/2/025002},
url = {https://dx.doi.org/10.1088/0264-9381/26/2/025002},
year = {2008},
month = {dec},
publisher = {},
volume = {26},
number = {2},
pages = {025002},
author = {Colosi, Daniele and Rovelli, Carlo},
title = {{What is a particle?}},
journal = {Classical and Quantum Gravity},
}

@article{Takagi1986,
  title="{Vacuum noise and stress induced by uniform acceleration: Hawking-Unruh effect in Rindler manifold of arbitrary dimension}",
  author={Takagi, Shin},
  journal={Progress of Theoretical Physics Supplement},
  volume={88},
  pages={1--142},
  year={1986},
  publisher={Oxford Academic}
}

@misc{Azizi2025KappaGamma,
  title={{Phase-Induced Particle Creation in the Kappa-Gamma Vacuum}}, 
  author={Arash Azizi},
  year={2025},
  eprint={2507.05299},
  archivePrefix={arXiv},
  primaryClass={hep-th},
  url={https://arxiv.org/abs/2507.05299}, 
}

@article{Azizi2025Tunable,
  title = {Tunable Unruh effect: Accelerated detectors in kappa-Rindler vacua},
  author = {Azizi, Arash},
  journal = {Phys. Rev. D},
  volume = {112},
  issue = {6},
  pages = {065018},
  numpages = {12},
  year = {2025},
  month = {Sep},
  publisher = {American Physical Society},
  doi = {10.1103/52lg-5sxr},
  url = {https://link.aps.org/doi/10.1103/52lg-5sxr}
}

@article{Azizi2025KappaPW,
  title = {{Kappa plane wave modes and continuous squeezing in quantum field theory}},
  author = {Azizi, Arash},
  journal = {Phys. Rev. D},
  volume = {112},
  issue = {2},
  pages = {025018},
  numpages = {15},
  year = {2025},
  month = {Jul},
  publisher = {American Physical Society},
  doi = {10.1103/ng6c-yvwm},
  url = {https://link.aps.org/doi/10.1103/ng6c-yvwm}
}

@article{Azizi2023JHEP,
  author       = {Arash Azizi},
  title        = {{Kappa vacua: enhancing the Unruh temperature}},
  journal      = {Journal of High Energy Physics},
  year         = {2023},
  volume       = {2023},
  number       = {7},
  pages        = {64},
  doi          = {10.1007/JHEP07(2023)064},
  url          = {https://doi.org/10.1007/JHEP07(2023)064},
  issn         = {1029-8479}
}

@misc{Azizi2022Kappashort,
      title={{Kappa vacua: Infinite number of new vacua in two-dimensional quantum field theory}}, 
      author={Arash Azizi},
      year={2023},
      eprint={2212.03781},
      archivePrefix={arXiv},
      primaryClass={hep-th},
      url={https://arxiv.org/abs/2212.03781}, 
}

@article{Martin_Schwinger1959,
  title = {{Theory of Many-Particle Systems. I}},
  author = {Martin, Paul C. and Schwinger, Julian},
  journal = {Phys. Rev.},
  volume = {115},
  issue = {6},
  pages = {1342--1373},
  numpages = {0},
  year = {1959},
  month = {Sep},
  publisher = {American Physical Society},
  doi = {10.1103/PhysRev.115.1342},
  url = {https://link.aps.org/doi/10.1103/PhysRev.115.1342}
}

@article{Kubo1957,
author = {Kubo ,Ryogo},
title = {{Statistical-Mechanical Theory of Irreversible Processes. I. General Theory and Simple Applications to Magnetic and Conduction Problems}},
journal = {Journal of the Physical Society of Japan},
volume = {12},
number = {6},
pages = {570-586},
year = {1957},
doi = {10.1143/JPSJ.12.570},
URL = {https://doi.org/10.1143/JPSJ.12.570},
}

@article{Moore1970,
    author = {Moore, Gerald T.},
    title = {{Quantum Theory of the Electromagnetic Field in a Variable‐Length One‐Dimensional Cavity}},
    journal = {Journal of Mathematical Physics},
    volume = {11},
    number = {9},
    pages = {2679-2691},
    year = {1970},
    month = {09},
    issn = {0022-2488},
    doi = {10.1063/1.1665432},
    url = {https://doi.org/10.1063/1.1665432},
   }

@article{Fulling_Davies1976,
author = {Fulling, S. A.  and Davies, P. C. W. },
title = {{Radiation from a moving mirror in two dimensional space-time: conformal anomaly}},
journal = {Proceedings of the Royal Society of London. A. Mathematical and Physical Sciences},
volume = {348},
number = {1654},
pages = {393-414},
year = {1976},
doi = {10.1098/rspa.1976.0045},
URL = {https://royalsocietypublishing.org/doi/abs/10.1098/rspa.1976.0045},
}

@article{Fulling_Davies1977,
author = {Davies, P. C. W.  and Fulling, S. A. },
title = {{Radiation from moving mirrors and from black holes}},
journal = {Proceedings of the Royal Society of London. A. Mathematical and Physical Sciences},
volume = {356},
number = {1685},
pages = {237-257},
year = {1977},
doi = {10.1098/rspa.1977.0130},
URL = {https://royalsocietypublishing.org/doi/abs/10.1098/rspa.1977.0130},
}

@article{Carlitz_Willey1987,
  title = {{Reflections on moving mirrors}},
  author = {Carlitz, Robert D. and Willey, Raymond S.},
  journal = {Phys. Rev. D},
  volume = {36},
  issue = {8},
  pages = {2327--2335},
  numpages = {0},
  year = {1987},
  month = {Oct},
  publisher = {American Physical Society},
  doi = {10.1103/PhysRevD.36.2327},
  url = {https://link.aps.org/doi/10.1103/PhysRevD.36.2327}
}

@article{Chen_Mourou2017PRL,
  title = {{Accelerating Plasma Mirrors to Investigate the Black Hole Information Loss Paradox}},
  author = {Chen, Pisin and Mourou, Gerard},
  journal = {Phys. Rev. Lett.},
  volume = {118},
  issue = {4},
  pages = {045001},
  numpages = {5},
  year = {2017},
  month = {Jan},
  publisher = {American Physical Society},
  doi = {10.1103/PhysRevLett.118.045001},
  url = {https://link.aps.org/doi/10.1103/PhysRevLett.118.045001}
}

@article{Svidzinsky2018PRL,
  title = {{Excitation of an Atom by a Uniformly Accelerated Mirror through Virtual Transitions}},
  author = {Svidzinsky, Anatoly A. and Ben-Benjamin, Jonathan S. and Fulling, Stephen A. and Page, Don N.},
  journal = {Phys. Rev. Lett.},
  volume = {121},
  issue = {7},
  pages = {071301},
  numpages = {6},
  year = {2018},
  month = {Aug},
  publisher = {American Physical Society},
  doi = {10.1103/PhysRevLett.121.071301},
  url = {https://link.aps.org/doi/10.1103/PhysRevLett.121.071301}
}

@article{Good2020Wilczek,
  title = {{Moving mirror model for quasithermal radiation fields}},
  author = {Good, Michael R. R. and Linder, Eric V. and Wilczek, Frank},
  journal = {Phys. Rev. D},
  volume = {101},
  issue = {2},
  pages = {025012},
  numpages = {6},
  year = {2020},
  month = {Jan},
  publisher = {American Physical Society},
  doi = {10.1103/PhysRevD.101.025012},
  url = {https://link.aps.org/doi/10.1103/PhysRevD.101.025012}
}

@article{Bisognano_Wichmann1975,
    author = {Bisognano, Joseph J. and Wichmann, Eyvind H.},
    title = {{On the duality condition for a Hermitian scalar field}},
    journal = {Journal of Mathematical Physics},
    volume = {16},
    number = {4},
    pages = {985-1007},
    year = {1975},
    month = {04},    
    issn = {0022-2488},
    doi = {10.1063/1.522605},
    url = {https://doi.org/10.1063/1.522605},
    
}

@article{Sewell1982,
title = {{Quantum fields on manifolds: PCT and gravitationally induced thermal states}},
journal = {Annals of Physics},
volume = {141},
number = {2},
pages = {201-224},
year = {1982},
issn = {0003-4916},
doi = {https://doi.org/10.1016/0003-4916(82)90285-8},
url = {https://www.sciencedirect.com/science/article/pii/0003491682902858},
author = {Geoffrey L Sewell},
}

@book{Gradshteyn_Ryzhik2014,
  author    = {Gradshteyn, I. S. and Ryzhik, I. M.},
  title     = {Table of Integrals, Series, and Products},
  edition   = {7th},
  publisher = {Academic Press},
  address   = {Amsterdam},
  year      = {2014},
  isbn      = {978-0-12-384933-5},
  DOI       = {https://doi.org/10.1016/C2010-0-64839-5}
}
\end{document}